\documentclass[conference]{IEEEtran}
\IEEEoverridecommandlockouts
\ifCLASSINFOpdf
\else
\fi
\usepackage{amsmath}
\usepackage{listings}
\usepackage{xcolor}
\usepackage{verbatim}
\usepackage{multirow}
\usepackage{graphicx}
\graphicspath{{./Figs/}}
\usepackage{url}
\usepackage{wasysym}
\usepackage{textcomp}
\usepackage{subcaption}
\usepackage{amssymb}
\usepackage{cite}
\usepackage{booktabs}
\usepackage{longtable}
\newcommand{\tabincell}[2]{\begin{tabular}{@{}#1@{}}#2\end{tabular}}
\usepackage{array}
\usepackage{rotating}
\newcommand{\PreserveBackslash}[1]{\let\temp=\\#1\let\\=\temp}
\newcolumntype{C}[1]{>{\PreserveBackslash\centering}p{#1}}
\newcolumntype{R}[1]{>{\PreserveBackslash\raggedleft}p{#1}}
\newcolumntype{L}[1]{>{\PreserveBackslash\raggedright}p{#1}}

\usepackage[linesnumbered,lined, noend, algoruled,norelsize]{algorithm2e}
\usepackage{pifont}
\usepackage{mathtools}
\usepackage[colorlinks,linkcolor=blue,anchorcolor=blue,citecolor=blue]{hyperref}
\usepackage{flushend}
\usepackage{ulem}
\usepackage{algorithm2e}
\usepackage{seqsplit}
\SetAlFnt{\small}
\SetAlCapFnt{\small}
\SetAlCapNameFnt{\small}
\begin{document}
%

\title{STAN: Towards Describing Bytecodes of \\Smart Contract}

\author{
	\IEEEauthorblockN{
		Xiaoqi Li\IEEEauthorrefmark{1},
		Ting Chen\IEEEauthorrefmark{2},
		Xiapu Luo\IEEEauthorrefmark{1}\thanks{\IEEEauthorrefmark{5} The corresponding author.}\IEEEauthorrefmark{5},
		Tao Zhang\IEEEauthorrefmark{3},
		Le Yu\IEEEauthorrefmark{1},
		Zhou Xu\IEEEauthorrefmark{1}\IEEEauthorrefmark{4}	
	}
	\IEEEauthorblockA{
		\IEEEauthorrefmark{1}Department of Computing, The Hong Kong Polytechnic University, Hong Kong SAR, China\\
		\IEEEauthorrefmark{2}Center for Cybersecurity, University of Electronic Science and Technology of China, Chengdu, China\\
		\IEEEauthorrefmark{3}Faculty of Information Technology, Macau University of Science and Technology, Macau SAR, China\\
		\IEEEauthorrefmark{4}School of Big Data and Software Engineering, Chongqing University, Chongqing, China\\
		Email: csxqli@gmail.com, brokendragon@uestc.edu.cn, csxluo@comp.polyu.edu.hk}
}

\maketitle

\begin{abstract}
\label{abs}
More than eight million smart contracts have been deployed into Ethereum, which is the most popular blockchain that supports smart contract. However, less than 1\% of deployed smart contracts are open-source, and it is difficult for users to understand the functionality and internal mechanism of those closed-source contracts. Although a few decompilers for smart contracts have been recently proposed, it is still not easy for users to grasp the semantic information of the contract, not to mention the potential misleading due to decompilation errors. In this paper, we propose the \textit{first} system named \textsc{Stan} to generate descriptions for the bytecodes of smart contracts to help users comprehend them. In particular, for each interface in a smart contract, \textsc{Stan} can generate four categories of descriptions, including functionality description, usage description, behavior description, and payment description, by leveraging symbolic execution and NLP (Natural Language Processing) techniques. Extensive experiments show that \textsc{Stan} can generate adequate, accurate and readable descriptions for contract's bytecodes, which have practical value for users.
\end{abstract}


\section{Introduction}
\vspace{1ex}

Since its inception, blockchain technology has shown promising application prospects from cryptocurrency to a variety of forms, such as medicine~\cite{yue2016healthcare}\cite{esposito2018blockchain} and cloud computing~\cite{samaniego2016blockchain}\cite{liang2017provchain}. As the program deployed and executed in blockchain, smart contract is the core technology in the 2.0 era of blockchain~\cite{zheng2018blockchain}. Through developing smart contracts, developers can realize rich logic and greatly expand the capabilities of blockchain system. As the most popular blockchain system that supports smart contract, Ethereum can complete one million transactions per day~\cite{2}. More than eight million smart contracts have already been deployed in Ethereum, while only less than 1\% are open-source~\cite{1_2}.

Unfortunately, facing the bytecodes of deployed smart contracts, it is difficult to quickly and comprehensively understand their details~\cite{grech2019gigahorse}\cite{zhou2018erays}, which leads to two issues. First, when users encounter a deployed contract, they usually do not know exactly how to use it, because users do not know the interfaces (i.e., external/public functions) of the bytecodes or the specific functionalities of its interfaces; Second, although some contracts are open-source, the blockchain system only stores the runtime bytecodes of them~\cite{7}. Usually common users cannot easily comprehend the contracts' sources published on websites (e.g., Etherscan~\cite{1}), not to mention the bytecodes of these contracts. Note that all the bytecodes mentioned in this paper refer to runtime bytecodes.

The root cause of above problems is the lack of tools to comprehensively summarize the functionalities of contract's bytecodes. Although a few decompilers for smart contracts have been recently proposed to turn contracts' bytecodes into user-defined IR (Intermediate Representation)~\cite{grech2019gigahorse} or Solidity sources~\cite{suiche2017porosity}, it is still not easy for users to grasp the semantic information of the contract, not to mention the potential misleading due to decompilation errors. Some other studies leverage symbolic execution~\cite{luu2016making}\cite{23}, static analysis~\cite{tikhomirov2018smartcheck}\cite{25}, dynamic analysis~\cite{CFI,TokenScope}, or formal methods~\cite{hirai2017defining}\cite{sergey2018temporal} to analyze smart contracts for detecting security issues, whose purposes are different from this paper. 

\begin{figure}[ht]
	\centering
	\vspace*{0ex}
	\includegraphics[width=3.40in]{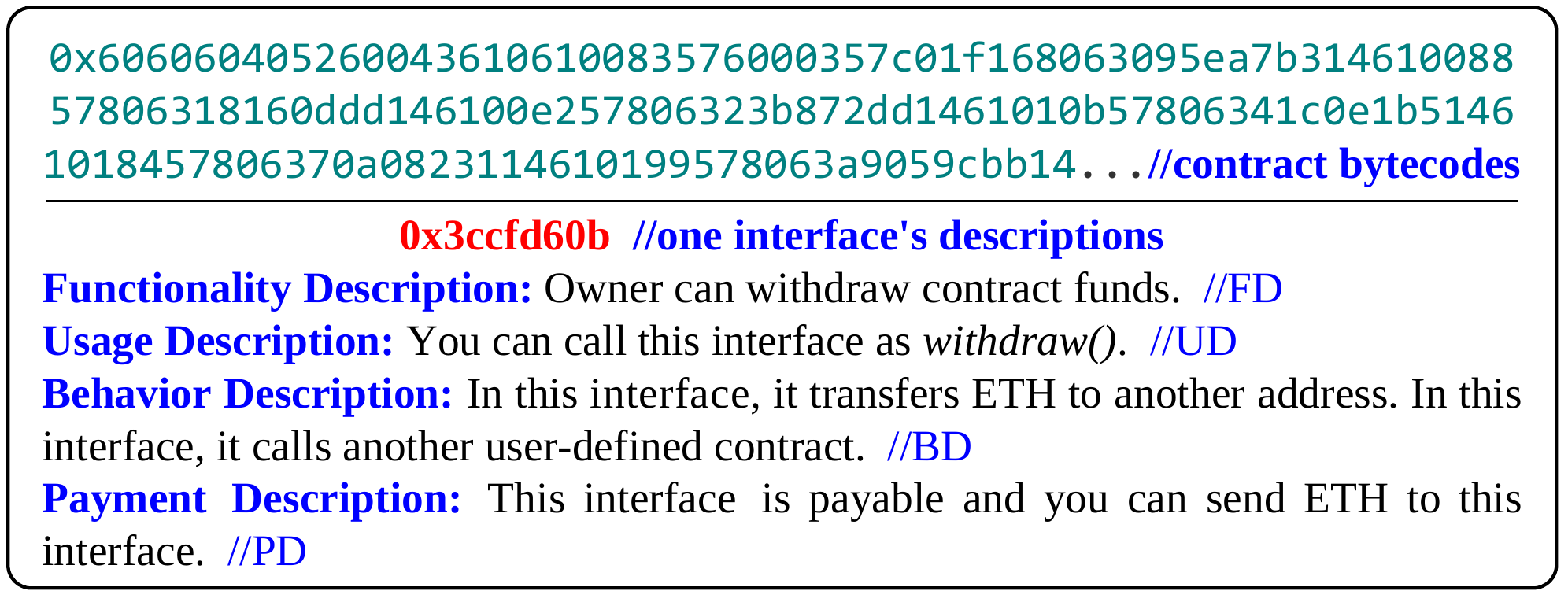}
	\vspace*{0ex}
	\caption{The runtime bytecodes of one closed-source contract (address at Mainnet: 0x68854ed29d6feca85242a9b5c00b9e93895a5403) and the descriptions for one interface generated through \textsc{Stan}.}
	\vspace{1ex}
	\label{des_example}
\end{figure}
In this paper, we propose and implement a system named \textsc{Stan} (deScribe byTecodes of smArt coNtract), which can analyze the runtime bytecodes of smart contract and automatically describe its interfaces in natural language, enabling users to quickly and thoroughly understand closed-source contracts. One motivating example of \textsc{Stan}'s descriptions for a smart contract is shown in Figure~\ref{des_example}. Given the address of target contract, \textsc{Stan} can automatically acquire its runtime bytecodes and describe every interface from four aspects. The functionality description summarizes the interface's functionality, and usage description tells the user how to call this interface. The behavior description describes message-call related behaviors within the interface, and payment description describes whether the interface can receive ETH.

The main contributions of this paper are as follows:

(1) To the best of our knowledge, we conduct the \textit{first} research of describing the bytecodes of smart contracts in natural language. For closed-source contract's bytecodes, \textsc{Stan} can generate four categories of descriptions for each interface.

(2) We leverage program analysis and NLP techniques to describe bytecodes. We analyze bytecodes through symbolic execution and generate readable descriptions following standard workflow of NLG (Natural Language Generation) system. 

(3) We evaluate the generated descriptions from three aspects. We develop a tool named \textsc{Scans}, which statically analyzes contract sources to evaluate descriptions' adequacy and accuracy. We also evaluate descriptions' readability through questionnaires and statistical methods.

\section{Background}
\label{file:background}
This section briefly introduces necessary background. 

\subsection{Ethereum}
Ethereum has two types of accounts, namely EOA (Externally Owned Account) and contract account~\cite{wan2019discussed}. Users can create EOAs and store ETH (native cryptocurrency in Ethereum). Users can send transactions using the private key associated to the EOA address, including ETH transfers and contract calls~\cite{chen2018understanding}. The contract account is created by EOA or another contract account. Besides ETH, the contract account contains the bytecodes and storage variables of smart contract.

\subsection{Smart Contract}
In Ethereum, each node runs an EVM (Ethereum Virtual Machine), and the bytecodes of contract are executed in EVM~\cite{zou2019}. Smart contract can be developed through several Turing complete languages, such as Solidity (the recommended language), Serpent, and Vyper~\cite{zou2019}. Therefore, smart contract can implement complex logics. 

\noindent\underline{Contract Interface:} It denotes functions that can be called externally by EOA or other deployed contract. \textit{``external''} or \textsl{``public''} functions can be invoked by others.
%
%
If a contract is open-source in Etherscan~\cite{1}, the most popular Ethereum block explorer, users can retrieve contract's ABI (Application Binary Interface) to get its interface information.

\noindent\underline{Contract Invocation:} After a smart contract is deployed to Ethereum, its interfaces can be called through transactions~\cite{chen2020understanding}. Gas is the basic unit of resource consumption for transactions in Ethereum. Invoking smart contracts through transactions requires a certain amount of gas\cite{chen2017adaptive}. When a smart contract is running in EVM, each opcode consumes 
some gas, whose value is defined in the Yellow Paper~\cite{7}.

\noindent\underline{Message-call:} There are two kinds of transactions in Ethereum, namely normal transaction (i.e., sent from EOA) and internal transaction (i.e., sent from contract). Through message-call, smart contract can interact with other EOAs or contract accounts, which typically cause internal transactions. One normal transaction may include several internal transactions, and message-call usually comes with the occurrence of sensitive behaviors. For example, through exploiting recursive message-call vulnerability, the criminals stole more than 60 million dollars from smart contract \textsc{Thedao}~\cite{li2020survey}. We analyze four different message-call related behaviors in this paper.

\noindent\underline{DevDoc:} Ethereum NatSpec (Natural Specification)~\cite{11} prescribes the writing specifications of DevDoc (Developer Documentation) in the contract's sources. For example, with the annotation field `details', contract developers can explain the functionality of the interface.

\noindent\underline{ERCDoc:} In EIP (Ethereum Improvement Proposal)~\cite{10}, there is ERCDoc (ERC Documentation), which prescribes standard interfaces for tokens in Ethereum. For example, ERC20 is the most popular token standard and there already exist more than 238,000 deployed ERC20 tokens~\cite{31}.

\begin{figure*}[ht]
	\centering
	\vspace*{0ex}
	\includegraphics[width=7.00in]{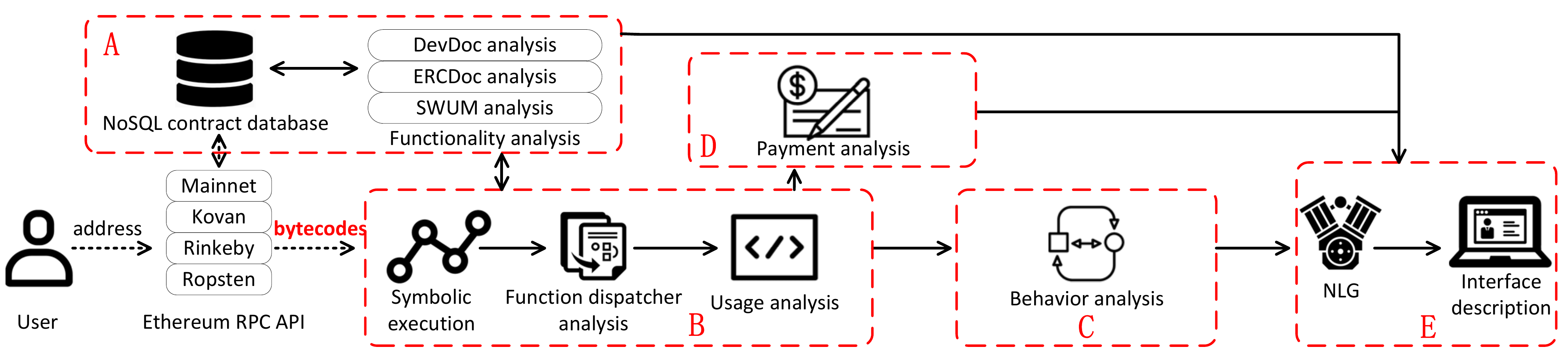}
	\vspace*{0ex}
	\caption{Overview of \textsc{Stan}'s architecture (best viewed in color). A is functionality analysis module; B is usage analysis module; C is behavior analysis module; D is payment analysis module; E is NLG module.}
	\vspace{0ex}
	\label{fig-architecture}
\end{figure*}

\section{\textsc{Stan} System}
\label{file:system}
The overview of \textsc{Stan}'s architecture is shown in Figure~\ref{fig-architecture}, which mainly consists of five modules: 

(1) Functionality analysis module (Section \ref{sec:general_module}): \textsc{Stan} conducts contract-oriented analysis through NLP techniques to generate functionality related phrases for interfaces. Note that the functionality of \textsc{Stan} is to describe closed-source contracts' bytecodes, and we analyze open-source contracts and metadata to provide help for bytecodes' analysis through discovering identical bytes signatures. 

(2) Usage analysis module (Section \ref{sec:usage_module}): \textsc{Stan} extracts function bytes signatures from function dispatcher and reverse them into corresponding text signatures. From text signature (e.g., $transfer(address,uint256)$), users can know function's name and parameter's configuration, which are used to call the interface. 

(3) Behavior analysis module (Section \ref{sec:behavior_module}): \textsc{Stan} analyzes external/public functions to generate intermediate information for message-call related behaviors leveraging symbolic execution. Through analyzing opcodes and operands, we recognize four kinds of sensitive message-call behaviors (e.g., ETH transfer, contract deployment, contract call). 

(4) Payment analysis module (Section \ref{sec:feature_module}): \textsc{Stan} analyzes external/public functions to generate intermediate information for payment feature through symbolic execution. We construct CFG (Control Flow Graph) to recognize two kinds of payment patterns, indicating whether the interface is payable. 

(5) NLG module (Section \ref{sec:nlg_module}): \textsc{Stan} generates the final readable interface descriptions leveraging the results of previous four modules. The NLG process follows the standard workflow of NLG system, i.e., document planner, micro-planner, and surface realizer.

\subsection{Functionality analysis module}
\label{sec:general_module}
\subsubsection{DevDoc and ERCDoc analysis}
\label{sec:devdoc}

In this section, we analyze DevDoc and ERCDoc to generate phrases that summarize interfaces' functionalities. The result is stored in \textsc{Stan}'s database and will be used to facilitate the describing of bytecodes. In the next section, we analyze interfaces without DevDoc or ERCDoc. 

Through \textsc{Scans}' static analysis for 129,737 open-source contracts, there are 5.36\% (6,954) sources with DevDoc. We show the process of DevDoc analysis through a function, whose text signature is `totalSupply()', which is divided into four steps. First, we leverage \textsc{Scans} to perform static analysis of sources and parse their DevDocs, to extract all the `details' annotations for functions with signature `totalSupply()'. 

Second, we aggregate the `details' annotations of functions, whose signatures are the same and appear in different contracts, into a single paragraph. Note that we only intercept the first sentence in each function instance's `details' annotations, as it is most closer to the goal of describing interfaces' functionalities. In addition, we pre-process the aggregated paragraph. In detail, we remove non-English sentences, identical sentences, meaningless special symbols, etc. After pre-processing, we obtain 53 different sentences, and all of them are written by the developer to describe the functionalities of `totalSupply()'. 

\vspace{0ex}
\begin{equation}
W(V_i)=(1-\overbrace{df}^{\mathclap{Damping~factor}})+df\times\sum_{V_j \in In(V_i)}{\frac{\overbrace{w_{ji}}^{\mathclap{Weight~of~E_{ji}}}}{\sum_{V_k \in Out(V_j)}{w_{jk}}}}W(V_j)
\label{textrank1}
\end{equation}
\vspace{-1ex}
\begin{tabbing}
	\hspace{1.2cm} \= \kill
	where: \> $In(V_i)$ is the set of vertices that point to $V_i$,\\
	\> $Out(V_j)$ is the set of vertices that $V_j$ points to.
\end{tabbing}

Third, we summarize the paragraph $T$ through TextRank Model. 
TextRank is a ranking model for natural language~\cite{mihalcea2004textrank} mainly used to unsupervised keywords extraction for texts. We conduct word segmentation (segmented by spaces), part-of-speech tagging on the paragraph, and filter out stop words. Then we build the keyword graph G = (V, E) through TextRank Model, whose vertice set is composed of word $t_{i}$. If two different $t_{i}$s appear in a window of length $k$, they have the co-occurrence relationship and there is an edge between the corresponding two vertices with specified weights. $V_i$'s weight is computed using Formula~\ref{textrank1}. We sort all the vertices according to their weights, to get several words with top weight values as keywords. At last, we extract key phrases (i.e., keywords with co-occurrence relationship) as summarization of the paragraph. 

\vspace{1ex}
\begin{equation}
Similarity(S_i,P_j) = \frac{|\{\overbrace{w_m}^{\mathclap{Words~in~sentence~and~phrase}}|w_m\in S_i \cap w_m\in P_j\}|}{|\{{w_n}|w_n\in S_i \cup w_n\in P_j\}|}
\label{textrank2}
\end{equation}
\vspace{-1ex}

\begin{table}[ht!]
	\centering
	\vspace*{0ex}
	\scriptsize
	\caption{Part of the statistics of ranked sentences in `details' paragraph for function signature `totalSupply()'. ID represents position ordinal of sentence in the `details' paragraph.}
	\label{ranked_sentences}
	\begin{tabular}{|c|c|c|c|}
		\hline
		\textbf{ID}&\textbf{MinHash Jaccard index}&\textbf{Sentence}\\ \hline
		$\star$47$\star$&$\star$0.183017870949962$\star$&$\star$`Total supply of tokens.'$\star$ \\ \hline
		36&0.161335751783794&`Returns the total token supply.'\\ \hline
		43&0.140755293788616&`Function to access total supply of tokens.' \\ \hline
		2&0.139558685183205&`Total Supply.' \\ \hline
		1&0.114068411467280&`Retrieves total supply.' \\ \hline
		40&0.091591782314505&`Obtain total number of tokens in existence.' \\ \hline
	\end{tabular}
	\vspace{1ex}
\end{table}

Fourth, to get the significance weight for different sentences in the paragraph, we calculate Jaccard index (shown in Formula~\ref{textrank2}) of each sentence $S_i$ to extracted key phrases $P_j$ through MinHash algorithm~\cite{broder1997resemblance}. At last, we sort the sentences according to their weight values, as shown in Table~\ref{ranked_sentences}, and select the highest weighted sentence (marked with $\star$) as the functionality phrase for the function with signature `totalSupply()'. 

Furthermore, we find that token-related function signatures have very high occurrence frequencies. Through \textsc{Scans}' static analysis for 129,737 open-source contracts, there are 67.56\% (87,652) sources with contract names that contain the keyword `erc'. In other words, approximately 67.56\% contracts implement the ERC standard token interfaces. Therefore, we get ERCDocs from EIPs and analyze standard token interfaces, to obtain token-related function signatures and their corresponding functionality phrases.

\begin{table}[ht!]
	\centering
	\vspace*{0ex}
	\scriptsize
	\caption{Quantity statistics of token interfaces defined in ERCDocs. $\star$ marks ERCDocs that extend ERC20. For example, ERC827 inherits 9 functions from ERC20 and defines 3 new functions. Note that the ERC1132 is also ERC20's extension; however, it only describes its 9 new functions. For all ERCDocs, we only extract external/public functions.}
	\label{tab_erc}
	\begin{tabular}{|c|c|c|c|}
		\hline
		\textbf{Documentation}&\textbf{Defined interfaces}&\textbf{Documentation} & \textbf{Defined interfaces}\\ \hline
		\textsc ERC20&9&ERC918&9 \\ \hline
		\textsc ERC721&17&ERC998&25\\ \hline
		\textsc ERC777&15&ERC1080&8 \\ \hline
		\textsc ERC$827{\star}$&\underline{9}+3&ERC$1132{\star}$&\underline{0}+9 \\ \hline
		\textsc ERC$884{\star}$&\underline{6}+11&ERC$1203{\star}$&\underline{6}+4 \\ \hline
		\textsc ERC900&10&ERC1410&12 \\ \hline
	\end{tabular}
	\vspace{1ex}
\end{table}

Because the writing structure of ERCDocs is not standardized or unified, it is difficult to parse their content automatically. First, we analyze the ERCDocs manually, extracting function signatures and their corresponding annotations defined in the documents. We have analyzed 12 popular token-related ERCDocs, and their relevant statistics are shown in Table~\ref{tab_erc}. Second, we combine different annotations of the same function signature into one paragraph. Third, we summarize functions' paragraphs through TextRank model separately, to generate functionality phrases for interfaces defined in ERCDocs. Eventually, we generate 115 different function signatures and their corresponding functionality phrases, which will be loaded into \textsc{Stan}'s database.

\subsubsection{SWUM analysis}
\label{sec:swum}

In this section, we generate functionality related phrases through SWUM for interfaces without DevDoc or ERCDoc, whose process is divided into three steps. 

SWUM (Software Word Usage Model) is used to extract linguistic information from program statements, including words in different parts of speech and the language relationship between them~\cite{sridhara2010towards}. We use some examples to interpret the process. First, for functions that follow standard naming conventions, we segment their text signatures through specific rules. We analyze three types of naming conventions, i.e., Camel case (e.g., `isPresaleReady()'), Pascal case (e.g., `GiveBlockReward()'), and Snake case (e.g., `claimed\_tokens()'). For those functions that do not follow standard naming conventions, we leverage Zipf's law~\cite{wordninja} to conduct word segmentation. After word segmentation, we tag the words in part-of-speech to get a set of nouns, verbs, and so on. For example, the function signature `isPresaleReady()' is segmented into (`is', `presale', `ready') and tagged as (`VBZ', `NP', `ADJP'). 

\begin{figure}[ht]
	\centering
	\vspace*{0ex}
	\includegraphics[width=3.50in]{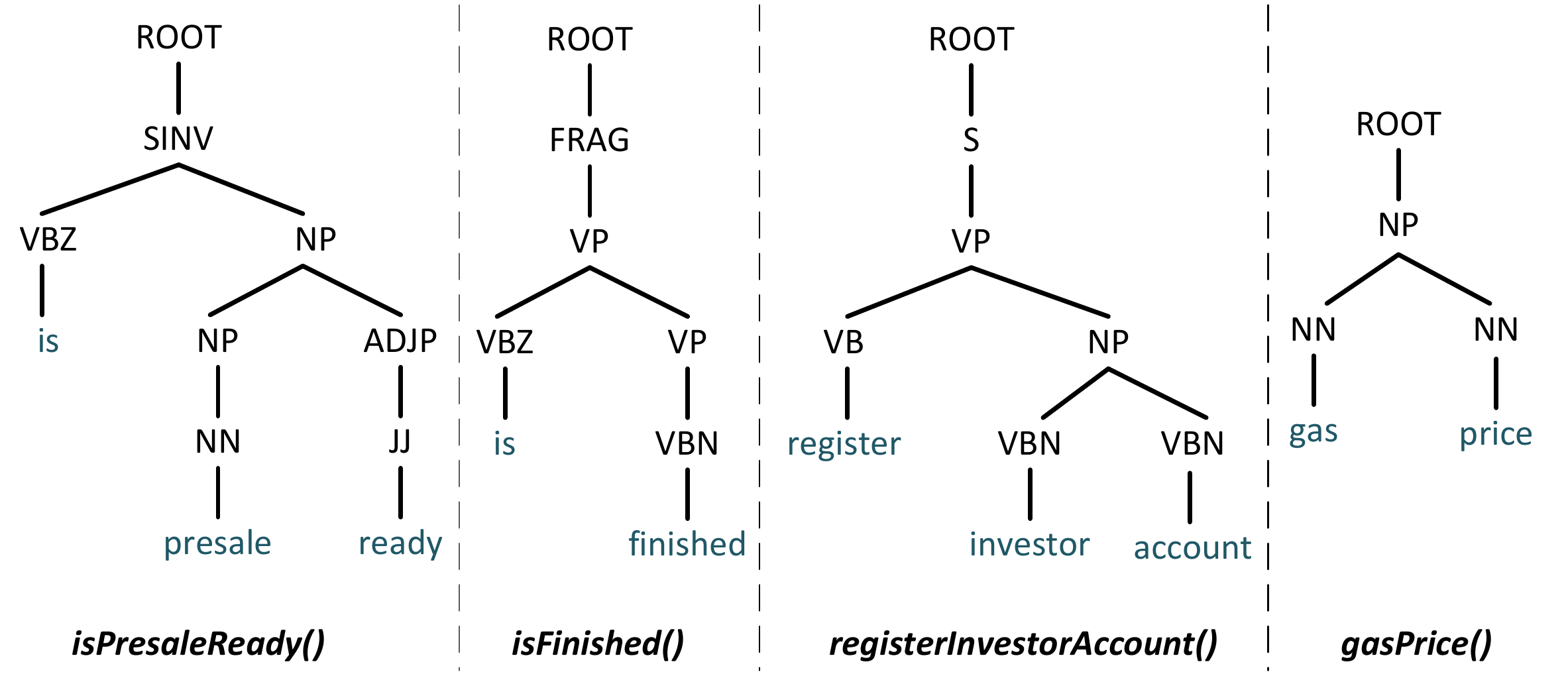}
	\vspace*{0ex}
	\caption{Four types and examples of syntax tree for function signature. SINV represents inverted declarative sentence; FRAG represents fragment; S represents simple declarative clause; NP represents noun phrase.}
	\vspace{0ex}
	\label{syntax_tree}
\end{figure}
Second, we analyze the linguistic relationship between the segmented words, such as subject-verb, verb-object, passive relations, etc. Leveraging Stanford parser~\cite{13}, we construct syntax tree of the text signature to analyze its linguistic relationship. We analyze four types of syntax tree, including SINV (inverted declarative sentence), FRAG (fragment), S (simple declarative clause), and NP (noun phrase). The structures and examples of these four syntax trees are shown in Figure~\ref{syntax_tree}. 

\begin{algorithm}[t] \caption{Phrase generation through syntax tree}
	\label{alg_syntax_tree}
	\begin{enumerate}   
		\item {\textbf{Input:} $S_f \leftarrow$ signature of function $f$}
		\item {$List_f \leftarrow$ WordSegmentation($S_f$)$~\rhd$\textcolor{blue}{through Camel, Pascal, Snake cases, or Zip'f law}}
		\item {$List_f \leftarrow$ Part-of-speech($List_f$)}
		\item {$ST_f \leftarrow$ StanfordParser($List_f$)$~\rhd$\textcolor{blue}{construct syntax tree}}
		\item {Switch(Type($ST_f$)):}
		\item {~~Case $SINV$:$~\rhd$\textcolor{blue}{inverted declarative sentence}}
		\item {~~~~$VE \leftarrow$ VerbSelector($ST_f$)}
		\item {~~~~$TP_f \leftarrow$ TemplateSelector($ST_f$)}
		\item {~~~~$P_f \leftarrow$ PhraseConstructor($List_f$, $ST_f$, $VE$, $TP_f$)}
		\item {~~Case $FRAG$:$~\rhd$\textcolor{blue}{fragment}}
		\item {~~~~$OB \leftarrow$ ObjectSelector($ST_f$)}
		\item {~~~~$TP_f \leftarrow$ TemplateSelector($ST_f$)}
		\item {~~~~$P_f \leftarrow$ PhraseConstructor($List_f$, $ST_f$, $OB$, $TP_f$)}
		\item {~~Case $S$:$~\rhd$\textcolor{blue}{simple declarative clause (no need to add new elements or select templates, because $List_f$ can be fully constructed into a phrase, and the sentence structure is fixed)}}
		\item {~~~~$P_f \leftarrow$ PhraseConstructor($List_f$, $ST_f$)}
		\item {~~Case $NP$:$~\rhd$\textcolor{blue}{noun phrase (no need to select templates, because the sentence structure is fixed)}}
		\item {~~~~$VE \leftarrow$ VerbSelector($ST_f$)}
		\item {~~~~$P_f \leftarrow$ PhraseConstructor($List_f$, $ST_f$, $VE$)}
		\item {\textbf{Output:} $P_f$}
	\end{enumerate} 
\end{algorithm}
\setlength{\textfloatsep}{9pt}
Third, the functionality phrase is generated using the analysis results from the previous two steps, and the process is shown in Algorithm~\ref{alg_syntax_tree}. In the algorithm of phrase generation, we analyze the structure of syntax tree, then select artificially designed verbs and templates to be assembled as phrases. For example, the function `isPresaleReady()' is classified as SINV type of syntax tree, and then we depth-first traverse the syntax tree, looking for the noun subject. Afterward, we select verb `check' and template \textit{`whether the NP VBZ ADJP'}, and generate the functionality phrase for this function as \textit{`Checks whether the presale is ready'}.
\subsubsection{Database construction}
\label{sec:db}

\begin{table}[ht!]
	\centering
	\vspace*{-0ex}
	\scriptsize
	\caption{Quantity statistics of deployed open-source smart contracts ($\star$ marks Testnets).}
	\label{sc_num}
	\begin{tabular}{|c|c|c|c|}
		\hline
		\textbf{Network name}&\textbf{Transactions}&\textbf{Block depth} & \textbf{Open-source contracts}\\ \hline
		\textsc Mainnet&506,822,407&8,234,086&\textbf{50,017} \\ \hline
		\textsc Kovan$\star$&23,994,109&12,485,019&\textbf{8,622}\\ \hline
		\textsc Rinkeby$\star$&38,609,478&4,807,975&\textbf{19,527} \\ \hline
		\textsc Ropsten$\star$&106,730,113&6,073,746&\textbf{51,571} \\ \hline
	\end{tabular}
	\vspace{-0ex}
\end{table}
We have crawled total of 129,737 deployed open-source contracts from Etherscan, whose statistics are shown in Table~\ref{sc_num}. Through DevDoc and ERCDoc analysis, we have generated 12,993 and 115 different function signatures and their corresponding functionality phrases respectively. Leveraging \textsc{Scans}, we totally extract 2,860,798 function text signatures from 129,737 sources' ABIs. As supplements, we obtain 147,724 from EFSD (Ethereum Function Signature Database)~\cite{3}, which is a public signature database that anyone can updates. Then we combine the text signatures extracted from ABIs and EFSD, removing duplicates, and use Keccak-256 hash algorithm to calculate bytes signature for each item. Eventually, we obtain 202,995 different text signatures and corresponding bytes signatures, which are used in SWUM analysis and usage analysis (Section~\ref{sec:usage_module}). We publish the above analysis results data on\textcolor{blue}{\url{https://figshare.com/articles/dataset/11650734}}. To the best of our knowledge, it is the most comprehensive Ethereum function signature public dataset. 

Note that we do not leverage code clone techniques in this paper because we only analyze open-source contract's function signature, not its function body. \textsc{Stan} can describe bytecodes that do not have corresponding sources.  The \textsc{Stan}'s database is used to reverse bytes signature and help to generate functionality and usage descriptions. Behavior and payment descriptions' generation does not use the database at all. We also use two different bytecodes datasets, one has corresponding sources and the other one does not have, to fully evaluate \textsc{Stan} in Section~\ref{file:evaluation}.

We construct contract database for \textsc{Stan}, and import these function-related data to assist in describing bytecodes of closed-source contracts. If one function's bytes signature in bytecodes can be retrieved in the database, we can use its related data to generate functionality and usage descriptions. We use MongoDB~\cite{14} to implement the database, and fully import the datasets published in the above URL into database to help describe bytecodes. When the final functionality descriptions are generated for one interface, we set specific priority rules to decide which field is used, which are presented in Section~\ref{sec:nlg_module}. The results of SWUM analysis are not loaded into database, because \textsc{Stan} directly generates descriptions from function signatures through SWUM if their DevDoc-related or ERCDoc-related phrases cannot be retrieved in database. Note that if one function's bytes signature in bytecodes cannot be retrieved in the database, its functionality and usage descriptions may not be generated properly.

\subsection{Usage analysis module}
\label{sec:usage_module}
In this section, we analyze the runtime bytecodes to recognize external functions' signatures, further to generate intermediate information for usage descriptions. 
\begin{figure}[ht]
	\centering
	\vspace*{0ex}
	\includegraphics[width=3.40in]{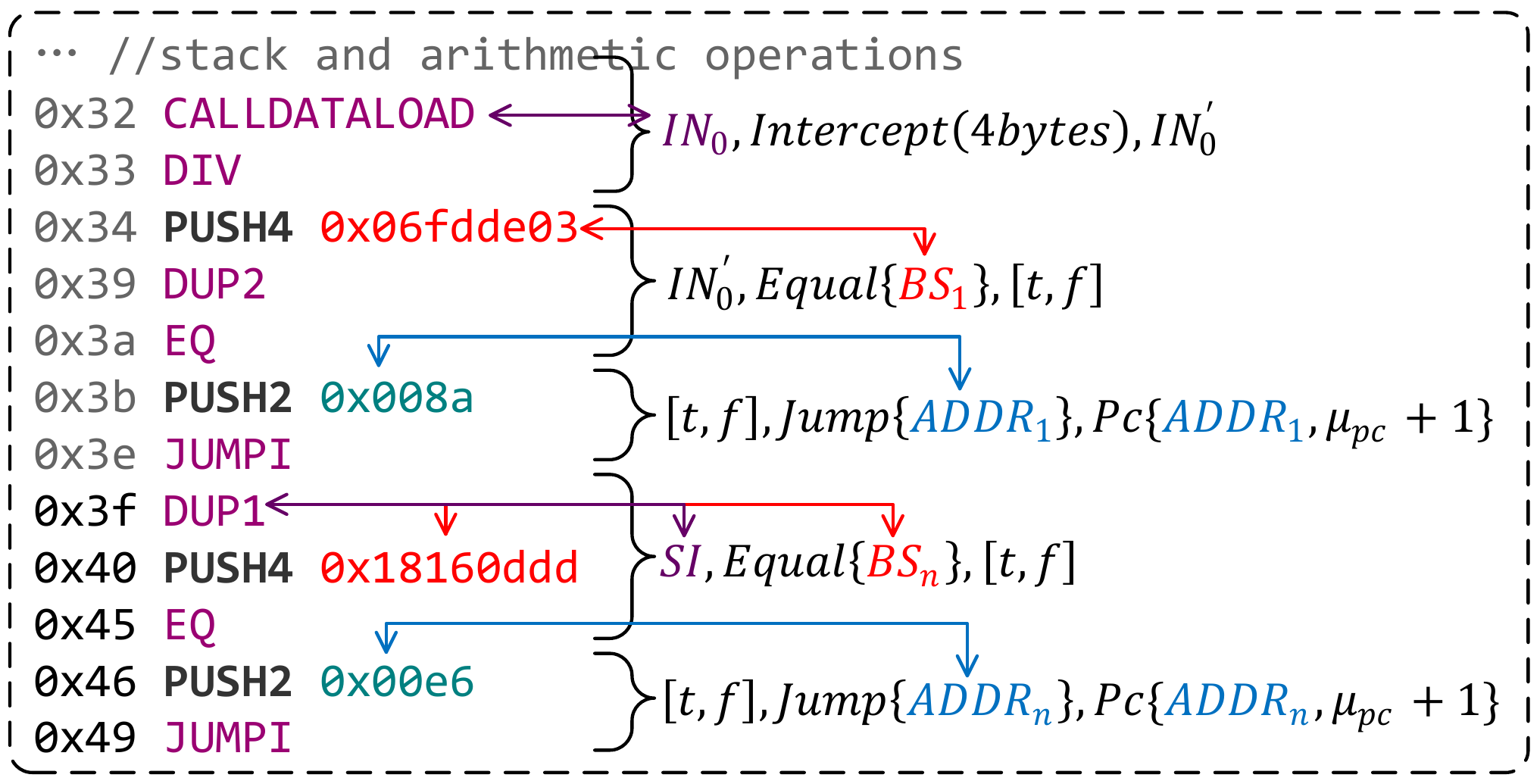}
	\vspace*{0ex}
	\caption{Bytecode snippet of two different types of function dispatcher (best viewed in color). Opcodes in program counter 0x32 to 0x3e belong to dispatcher type one, and opcodes in counter 0x3f to 0x49 belong to dispatcher type two.}
	\vspace{1ex}
	\label{dispatcher_example}
\end{figure}

The usage analysis is divided into three steps. First, leveraging \textsc{Oyente}~\cite{luu2016making}, which is a symbolic execution engine, we construct CFG of the runtime bytecodes. Second, we recognize function dispatchers in the CFG. The function dispatcher is used to compare the bytes signature encoded in transaction parameter with signatures in runtime bytecodes, to decide which function to execute exactly. There exist two different types of function dispatchers in bytecodes, and we use the bytecode snippet of one closed-source contract (Address at Mainnet: \seqsplit{0x50e57ada51fa82b5a3de6ebae3d21f88c8d3a672}) (shown in Figure~\ref{dispatcher_example}) to interpret their patterns. For the first type, which is usually located in the opcode block of initial entrance, it reads the first 32 bytes of the transaction's input data as $IN_0$. After intercepting the first 4 bytes of $IN_0$ into $IN_0^{'}$, it compares $IN_0^{'}$ with the first bytes signature in bytecodes $BS_1$. If $IN_0^{'}$ equals $BS_1$, the program counter will be changed to $ADDR_1$, which is the opcode block corresponding to function $BS_1$. Otherwise, the program counter will be changed to $\mu_{pc}+1$. For the second type of function dispatcher, it reads $SI$, which is the bytes signature of the transaction's target function, from the stack directly. Then it compares $SI$ with one of bytes signature in bytecodes $BS_{n}$. If $SI$ equals $BS_{n}$, the program counter will be changed to $ADDR_n$. Otherwise, the program counter will be changed to $\mu_{pc}+1$. 

Third, we extract function bytes signatures from function dispatchers (i.e., $(BS_1, 0x06fdde03)$ and $(BS_n, 0x18160ddd)$) and retrieve their corresponding text signatures through the contract database. Note that it may fail to retrieve text signatures, whose adequacy is evaluated in Section~\ref{rq1_adequacy}. Eventually, the extracted function bytes signatures and their corresponding text signatures (i.e., $(0x06fdde03, name()$ and $(0x18160ddd, totalSupply()$) act as intermediate information to be transferred to the NLG module (in Section~\ref{sec:nlg_module}) to generate usage descriptions.

\subsection{Behavior analysis module}
\label{sec:behavior_module}
In this section, we analyze four kinds of message-call behaviors in interface, further to generate intermediate information for behavior descriptions. The behavior analysis is divided into two steps. First, for every execution path in function body, we record the occurrence of message-call related opcodes and their corresponding operands through symbolic execution. Second, we analyze the recorded information of message-call related opcodes and operands, and summarize it into four different categories of interface behaviors listed in Table~\ref{mcall_table1}. 
\begin{table}[ht!]
	\centering
	\vspace{1ex}
	\scriptsize
	\caption{Four different categories of interface behaviors through message-call, and their corresponding opcodes and operands to be analyzed. $\star$ marks the behaviors that cause internal transactions.}
	\vspace{0ex}
	\label{mcall_table1}
	\begin{tabular}{|c|c|c|}
		\hline
		\textbf{Interface behavior}&\textbf{Analyzed opcode}&\textbf{Analyzed operand}\\ \hline
		\multirow{2}*{ETH transfer$\star$}&{CALL, CALLCODE}&{$P_v$}\\ \cline{2-3}
		~&{SELFDESTRUCT}&{\ding{55}}\\ \hline
		{Pre-compiled contract call}&{CALL}&{$P_a$}\\ \hline
		\multirow{2}*{User-defined contract call$\star$}&{CALL}&{$P_a$}\\ \cline{2-3}
		~&{\tabincell{c}{CALLCODE, STATICCALL,\\DELEGATECALL}}&{\ding{55}}\\ \hline
		{Contract deployment$\star$}&{CREATE}&{\ding{55}}\\ \hline
	\end{tabular}
	\vspace{2ex}
\end{table}

For ETH transfer behavior, there are two scenarios. In the first scenario, \texttt{CALL} or \texttt{CALLCODE} is bound to appear during function body execution. Further, we analyze their value field operand $P_v$, to check whether $P_v$ is a non-zero constant or symbolic value. If $P_v$ is a non-zero constant, it indicates the existence of a fixed amount of ETH transfer. If $P_v$ is a symbolic value, it indicates the existence of a non-fixed amount of ETH transfer, whose specific value is determined by function's input parameter or contract's storage. In the second scenario, \texttt{SELFDESTRUCT} is bound to appear and we do not need to analyze its operand. The occurrence of \texttt{SELFDESTRUCT} indicates that there is ETH transfer during contract's self-destruction. For PRE contract call behavior, \texttt{CALL} is bound to appear, and we analyze its target address field operand $P_a$, to check whether $P_a$ is a constant value from \textit{0x1} to \textit{0x8}. From version Metropolis~\cite{7}, Ethereum implements eight different PRE contracts. 

For user-defined contract call behavior, there are two scenarios. In the first scenario, \texttt{CALL} is bound to appear during function body execution, and its operand $P_a$ is not any of the addresses of PRE contracts. In the second scenario, \texttt{CALLCODE} or \texttt{STATICCALL} or \texttt{DELEGATECALL} is bound to appear during function body execution. We do not need to analyze their operands in this scenario, and the presence of any of them can prove the existence of user-defined contract call behavior. 

For contract deployment behavior, \texttt{CREATE} is bound to appear during function body execution, and we do not need to analyze its operands. The occurrence of \texttt{CREATE} indicates that there is inline assembly or call to its constructor in the function body, to deploy new contracts. At last, the target interface's specific message-call behavior category will act as intermediate information to be transferred to the NLG module to generate behavior descriptions.

\subsection{Payment analysis module}
\label{sec:feature_module}
In this section, we analyze whether the target interface is ETH payable, further to generate intermediate information for payment feature descriptions. When we develop smart contract with Solidity, a function needs to be decorated with modifier \textit{payable} in order to receive ETH through transactions. If users call a non-payable interface with ETH, the transaction execution will fail and waste user's gas. 
\begin{figure}[ht]
	\centering
	\vspace*{0ex}
	\includegraphics[width=3.40in]{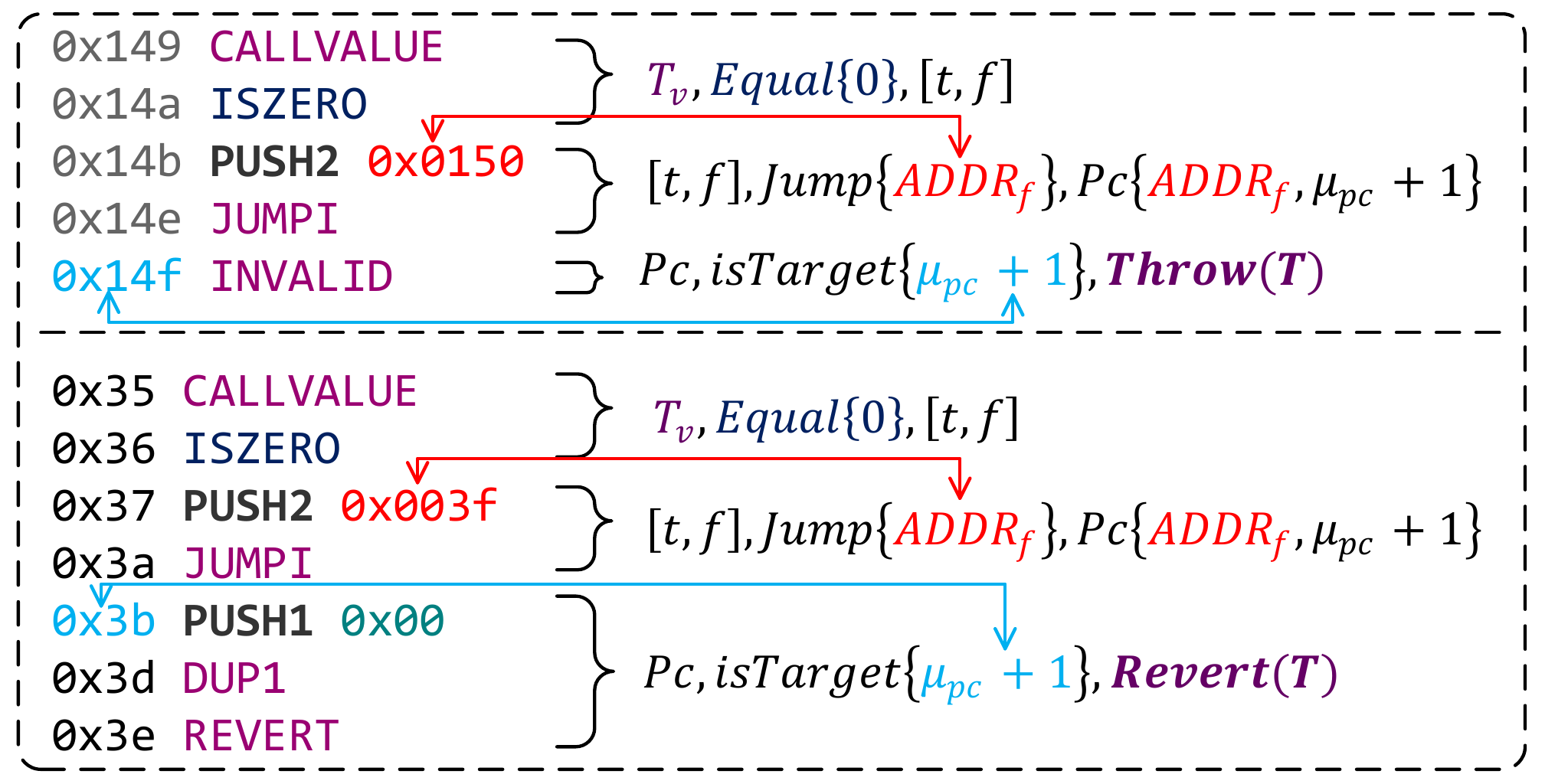}
	\vspace*{0ex}
	\caption{Bytecode snippet of two different types of payment operations (best viewed in color). Opcodes in program counter 0x149 to 0x14f (in function $0x66117276$) belong to non-payable type one, and opcodes in counter 0x35 to 0x3e (in function $0x62c06767$) belong to non-payable type two.}
	\vspace{1ex}
	\label{nonpayable_example}
\end{figure}

We recognize payment operations' patterns in the CFG of the target function body, to detect whether the interface is non-payable. We use two different bytecode snippets (contract A at Mainnet: \seqsplit{0x6ab6aac6a6f844e322a6c42b3185e1bc4cf56e42}, and B at Mainnet: \seqsplit{0xa3ed88f7c9bf7df33b7549bb8c5a889b6049504c}) to interpret payment patterns as shown in Figure~\ref{nonpayable_example}. 

It acquires transaction's value field $T_v$, and determine whether $T_v$ equals 0. If $T_v$ equals 0, the program counter will be changed to $ADDR_f$, which is the initial execution block's address of behaviors within function. If $T_v$ is not equal to 0, the program counter is changed to $\mu_{pc}+1$, which is the address of next execution block. In the execution block that started from $\mu_{pc}+1$, it throws or reverts the transaction due to different compiler \textsc{Solc}~\cite{29} versions. Before \textsc{Solc} version 0.4.12, it throws the transaction through \texttt{INVALID} (\textit{0xfe} in bytes). In this scenario, it rolls back all state changes and consumes the remaining gas. For example, one transaction (Hash at Rinkeby: \seqsplit{0x128907301beff5c56af8234e3c925567352696defffc89e453313e33ae73ac5d}) failed with \textit{invalid opcode error}, and it consumed all the 4,707,786 gas from the user. After \textsc{Solc} version 0.4.12 (including v0.4.12), it reverts the transaction through \texttt{REVERT} (\textit{0xfd} in bytes). In this scenario, it rolls back all state changes and returns the remaining gas to the user. For example, one transaction (Hash at Mainnet: \seqsplit{0x037a08b19bc3255e2feca42f6e08294ab8f7daa26e400d998b2a1368159216f2}) failed with \textit{inverted error}, and it consumes 22.53\% (22,525/100,000) of the gas given by the user. Therefore, in both scenarios, transaction will fail and cause the user's gas wasted. Note that either of our detected pattern's occurrence indicates that the target interface is non-payable. 

At last, the result of interfaces' analysis, \seqsplit{$(0x66117276, [Nonpayable, True])$} and \seqsplit{$(0x62c06767, [Nonpayable, True])$}, act as intermediate information to be transfered to the NLG module respectively to generate feature descriptions.

\subsection{NLG module}
\label{sec:nlg_module}
In this section, we generate the final readable interface descriptions leveraging the results of previous four analysis modules. The NLG module mainly consists of three steps, which are shown in Algorithm~\ref{alg_nlg_engine}.

\begin{algorithm}[t] \caption{Description generation through NLG module}
	\label{alg_nlg_engine}
	\begin{enumerate}   
		\item {\textbf{Input:} $IF_{f,u,b,p}$ $~\rhd$\textcolor{blue}{intermediate information of functionality, usage, behavior, and payment descriptions}}
		\item {\textcolor{blue}{\textit{\textbf{Step 1:} document planner}}}
		\item {$WIF \leftarrow$ WeightAssign($IF_{f,u,b,p}$)}
		\item {For $Element$ in $IF_{f,u,b,p}$:}
		\item {~~$WIF_{f,u,b,p} \leftarrow$ WeightAssign($Element$)}
		\item {\textcolor{blue}{\textit{\textbf{Step 2:} micro-planner}}}
		\item {For $P_{d,s,e}$ in $IF_f$:}
		\item {~~$C_{G_{d,s,e}}  \leftarrow$ NLGFactory-CreateClause($P_{d,s,e}$)}
		\item {For $Element$ in $IF_{u,b,p}$:}
		\item {~~$TP_{u,b,p}  \leftarrow$ TemplateSelector(Type($IF_{u,b,p}$))}
		\item {~~$C^{''}_{U,B,P} \leftarrow$ NLGFactory-CreateClause($TP_{u,b,p}$,$Element$)}
		\item {$C^{'}_{U,B,P} \leftarrow$ Aggregator($C^{''}_{U,B,P}$)}
		\item {\textcolor{blue}{\textit{\textbf{Step 3:} surface realizer}}}
		\item {$C_F \leftarrow$ WeightHighest($C_{G_{d,s,e}}$,$WIF_{g}$)}
		\item {$C_{U,B,P} \leftarrow$ WeightSort($C^{'}_{U,B,P}$,$WIF_{u,b,p}$)}
		\item {$D_{F,U,B,P} \leftarrow$ NLGFactory-CreateParagraph($C_{F,U,B,P}$)}
		\item {$D_i \leftarrow$ GrammarChecker(WeightSort($D_{F,U,B,P}$,$WIF$))}
		\item {\textbf{Output:} $D_i$}
	\end{enumerate} 
	\vspace{1ex}
\end{algorithm}

The first step is document planner for content determination and document structuring. After importing four categories of intermediate information $IF_{f,u,b,p}$ for the target interface, we give them four different weights to determine the order in which the descriptions appear. Note that specific weight values of $WIF$ are set depending on the degree of $IF$'s importance, which will be used to determine paragraphs' order. For example, we give $IF_f$ the highest weight and functionality description will appear first among the four kinds of descriptions. Similarly, we traverse specific elements in $IF_{f,u,b,p}$ and weight them, which will be used to select elements to generate sentences, and to determine the order of sentences within paragraphs. 

The second step is micro-planner for lexicalization and aggregation. Through $IF_f$, we parse three different kinds of functionality phrases $P_{d,s,e}$, which represent DevDoc-based, SWUM-based, and ERCDoc-based phrases. Leveraging NLGFactory APIs in \textsc{Simplenlg}~\cite{17}, which is a package used for language generation, we create complete sentences $C_{G_{d,s,e}}$ of functionality descriptions from $P_{d,s,e}$. Then we traverse specific elements in $IF_{u,b,p}$ and select sentence template $TP_{u,b,p}$ according to the category of $IF_{u,b,p}$. Using specific $TP_{u,b,p}$ and $Element$, we create complete sentences for usage, behavior, and payment descriptions $C^{''}_{U,B,P}$. Because there might exist sentences that are highly similar or identical in $C^{''}_{U,B,P}$, we set rules to aggregate these sentences. For example, when there are two ETH transfers in the same $IF_b$, we describe them only once. 

The third step is surface realizer for linguistic and structure realization. For the three kinds of sentences $C_{G_{d,s,e}}$, we select the highest weighted sentence to act as the interface's functionality description. In NLG module's implementation, we set the highest weights for ERCDoc-based sentences because their $IF_f$ are artificially analyzed and extracted from EIPs in Section~\ref{sec:devdoc}, which are more accurate than the other two kinds of sentences. For the sentences in $C^{'}_{U,B,P}$, we determine their appearance order according to their weight values in $WIF_{u,b,p}$. Then we create four different paragraphs $D_{F,U,B,P}$ from the four kinds of sentences $C_{F,U,B,P}$, leveraging NLGFactory APIs. After we adjust paragraphs' order of $D_{F,U,B,P}$ according to their weight values in $WIF$, we check their language grammar through \textsc{Languagecheck}~\cite{18}. At last, $D_i$ is output as the final descriptions for the target interface.

\section{Evaluation}
\label{file:evaluation}

We conduct a series of experiments to evaluate \textsc{Stan}, which are used for answering the following research questions: 

\textsl{RQ1 Adequacy:} How many contracts' bytecodes can be successfully described through \textsc{Stan}?

\textsl{RQ2 Accuracy:} To what extent can \textsc{Stan} accurately describe contracts' bytecodes?

\textsl{RQ3 Readability:} How is the readability of the generated descriptions for users?

\subsection{Datasets and Experimental Overview}
\label{sec:expoverview}
\begin{table}[ht!]
	\centering
	\vspace*{1ex}
	\scriptsize
	\caption{Quantity statistics of two kinds of contract bytecodes' datasets for evaluation.}
	\vspace{0ex}
	\label{st_dataset}
	\begin{tabular}{|c|c|c|c|c|c|}
		\hline
		\textbf{DS}&\textbf{Network}&\textbf{Bytecode}&\textbf{Destructed}&\textbf{Identical}&\textbf{Analyzed}\\ \hline
		\textbf{1}&Mainnet&6,920,465&N/A&6,803,635&116,830 \\ \hline
		\multirow{4}*{\textbf{2}}&Mainnet&50,017&725&1,398&47,894 \\ \cline{2-6}
		~&Kovan&8,622&79&514&8,029 \\ \cline{2-6}
		~&Rinkeby&19,527&228&269&19,030 \\ \cline{2-6}
		~&Ropsten&51,571&626&998&49,947 \\ \hline
	\end{tabular}
	\vspace{2ex}
\end{table}

In order to fully evaluate \textsc{Stan}, we create two kinds of bytecodes' datasets, which are shown in Table~\ref{st_dataset}. For DS1, we crawl 42,115,551 different accounts' information in 28 days from Mainnet. We resolve all crawled accounts, with 6,920,465 accounts containing bytecodes, indicating that these are contract accounts and they are not self-destructed. After deleting all accounts containing duplicate bytecodes, we obtain 116,830 different runtime bytecodes. For DS2, we crawl open-source contracts' bytecodes from Mainnet and three public Testnets. Then we delete accounts that contain empty bytecodes, which means that they are already self-destructed, and accounts containing duplicate bytecodes. \textbf{All the bytecodes in DS2 can retrieve corresponding source codes through Etherscan, which can facilitate us to evaluate the accuracy and readability of their descriptions generated from bytecodes.} The statistics also show that only less than 1\% (50,017/6,920,465) contracts are open-source. 

Considering that the symbolic execution consumes time and hardware resources, we run experiments through 4 cloud instances. These instances are all configured with Intel Xeon E312x 2.60GHz CPU and 8G RAM, running 64-bit Ubuntu 18.04. We randomly extract 800 runtime bytecodes from DS1 to constitute DS1', and 200 from each network (total of 800) in DS2 to constitute DS2'. \textbf{We have checked all the bytecodes in DS1' and they are not verified with source codes in Etherscan.} 

Before the evaluation, we run \textsc{Stan} to generate descriptions for DS1' and DS2', which include a total of 1,600 contracts' bytecodes. As a first step, we run the \textsc{Oyente}~\cite{luu2016making} engine on the datasets alone, in 25 hours, to remove those contracts that encounter timeout exception. There are 651 contracts' bytecodes, 357 in DS1' and 294 in DS2', executed without timeout. 
Second, we generate descriptions for these 651 contracts' bytecodes through \textsc{Stan}, with an average analysis time of 87.4s (including symbolic execution) per contract. 

\subsection{RQ1 Adequacy}
\label{rq1_adequacy}
\begin{table}[ht!]
	\centering
	\vspace*{1ex}
	\scriptsize
	\caption{Quantity statistics of the success rate of normally describing or tagging bytecodes. Note that we tag two kinds of insecure contracts, i.e., NF (No Function) contracts, and JE (Jump Exception) contracts.}
	\vspace{0ex}
	\label{st_adequacy1}
	\begin{tabular}{|c|c|c|c|c|c|}
		\hline
		\textbf{DS}&\textbf{Bytecodes}&\textbf{Described}&\textbf{NF tagged}&\textbf{JE tagged}&\textbf{Success rate}\\ \hline
		\textbf{1'}&357&292 (81.8\%)&62 (17.4\%)&3 (0.8\%)&100\% \\ \hline
		\textbf{2'}&294&294 (100\%)&0&0&100\% \\ \hline
	\end{tabular}
	\vspace{2ex}
\end{table}
In this section, we evaluate how many bytecodes can be successfully described through \textsc{Stan}. The quantity statistics of the success rate of normally describing or tagging bytecodes are shown in Table~\ref{st_adequacy1}. 

For DS1', there are 292 (81.8\%) bytecodes described normally through our NLG module. The other 65 bytecodes are tagged as insecure contracts, i.e., NF (No Function) contracts or JE (Jump Exception) contracts, which are advised not to be called. The NF contract has no external/public function and executes the same opcode snippet for each invocation. For example, one tagged NF contract (Address at Mainnet: \seqsplit{0x5170E3C93df0605F3b02b00d8C3D9a7235fcD1Ef}) is a honeypot contract, and it executes the same useless operations for each invocation. It wastes users' gas and can maliciously receive users' ETH attached in the transaction. Therefore, we tag this contract's bytecodes as \textit{``ALERT: This is an insecure NF contract!''} The JE contract has invalid jump destination(s) in its opcodes, which may encounter jump exception and exhaust users' gas. For one tagged JE contract (Address at Mainnet: \seqsplit{0x6a5dffaAdBCbeF3359a017cc5100908630364aBF}), regardless of which interface is called, it will encounter a runtime exception, which exhausts users' gas. Therefore, we tag this contract's bytecodes as \textit{``ALERT: This is an insecure JE contract!''} 

For DS2', all the 294 (100\%) contracts' bytecodes are normally described and no bytecodes tagged as insecure contracts. We can conclude that contracts with corresponding verified source codes are generally more secure.

\begin{table}[ht!]
	\centering
	\vspace*{1ex}
	\scriptsize
	\caption{Quantity statistics of interfaces described normally, and the success rate of four kinds of descriptions.}
	\vspace{-0ex}
	\label{st_adequacy2}
	\begin{tabular}{|c|c|c|c|c|}
		\hline
		\textbf{DS}&\textbf{Interfaces}&\textbf{FD}&\textbf{UD}&\textbf{BD \& PD}\\ \hline
		\textbf{1'}&3,179 (N/A)&{2,231 (70.2\%)}&{2,979 (93.7\%)}&{3,179 (100\%)} \\ \hline
		\textbf{2'}&4,180 (100\%)&{3,023 (72.3\%)}&{4,180 (100\%)}&{4,180 (100\%)} \\ \hline
	\end{tabular}
	\vspace{1ex}
\end{table}
We further evaluate \textsc{Stan}'s adequacy from the level of interfaces, whose statistics are shown in Table~\ref{st_adequacy2}. For the described 292 contracts' bytecodes in DS1', 3,179 interfaces are analyzed. There are 2,231 (70.2\%) interfaces' functionalities successfully described. Some interfaces are failed to generate functionality description because they are not included in DevDoc and ERCDoc analysis. In the meanwhile, they cannot be analyzed perfectly through SWUM, which are mainly reflected in two aspects. First, some functions are highly irregularly named. For example, function \texttt{caps(address)} cannot be recognized through Stanford parser~\cite{13} because ``caps'' is not a complete word or standard abbreviation. Second, some functions' syntax structure cannot be analyzed. For example, function \texttt{MAX\_INVESTMENTS\_BEFORE\_CHANGE()} cannot be classified into the four syntax trees we detect. There are 2,979 (93.7\%) text signatures recognized through our usage analysis and contract database. To the best of our knowledge, we already construct the most comprehensive function signature dataset. For the other 6.3\% functions, there is currently no viable way to identify their text signatures. 

For the 294 contracts' bytecodes in DS2', by using \textsc{Scans}, we acquire their corresponding source codes and totally extract 4,180 external/public functions statically. As shown in Table~\ref{st_adequacy2}, 100\% of these interfaces are analyzed and identified text signatures through usage analysis. Similar to DS1', 72.3\% interfaces' functionalities successfully described.

\textbf{Answer to RQ1 (Adequacy):} \textsc{Stan} can successfully describe or tag 100\% of the bytecodes in two datasets. Furthermore, 100\% of interfaces can be successfully described by two description modules (i.e., BD, PD). More than 93.7\% of interfaces' usage descriptions can be successfully generated, and more than 70.2\% of interfaces' functionalities can be successfully described. \textsc{Stan} can adequately describe bytecodes of smart contracts.

\begin{figure*}[htbp]
	\centering
	\begin{minipage}[t]{0.48\textwidth}
		\centering
		\includegraphics[width=3.40in]{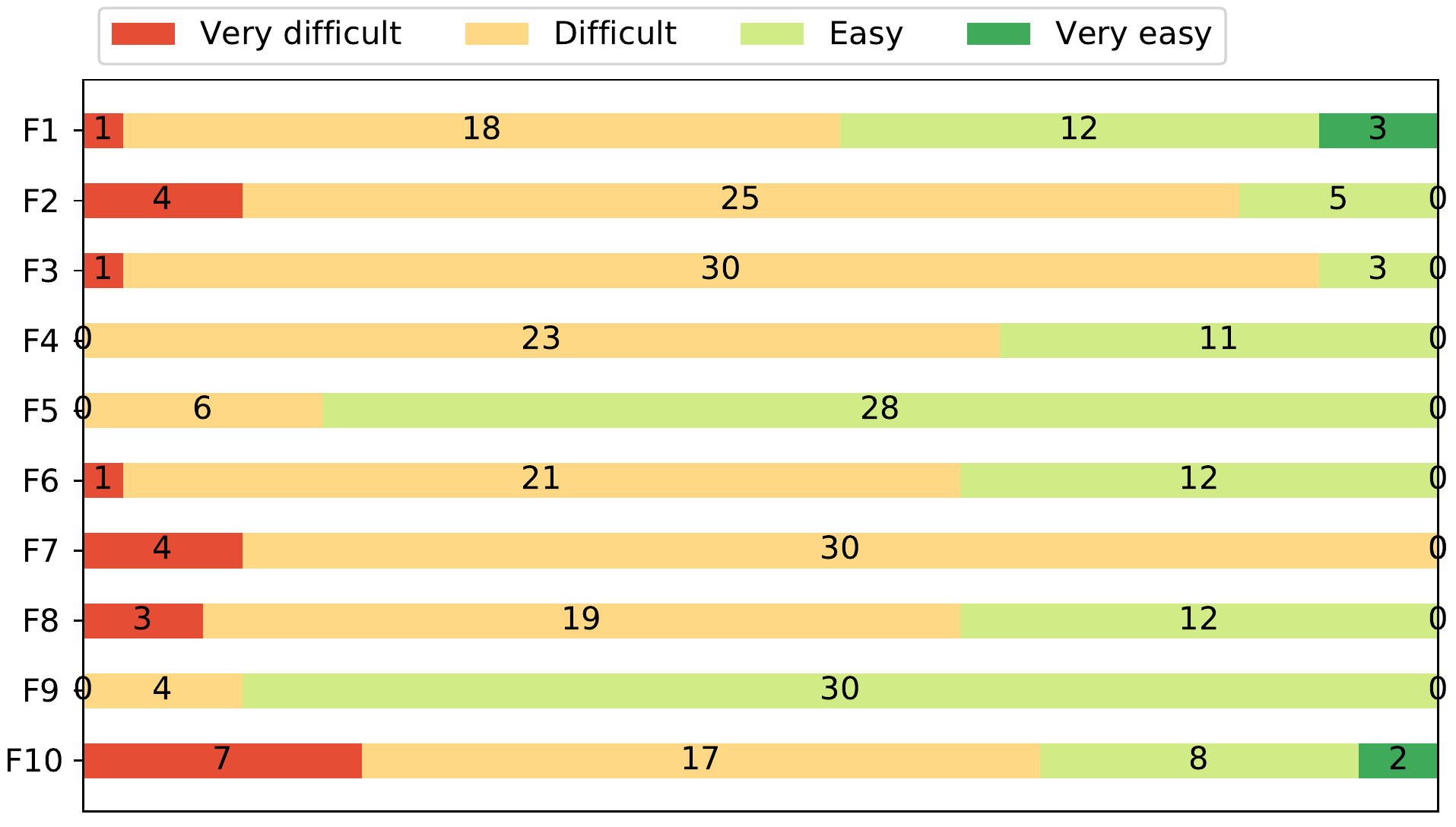}
		\caption{Quantity statistics of readability scores for 10 interfaces' annotations, which are written by developers in their source codes.}
		\label{fig_rq3_1}
		\vspace{-1ex}
	\end{minipage}
	\hspace{3ex}
	\begin{minipage}[t]{0.48\textwidth}
		\centering
		\includegraphics[width=3.40in]{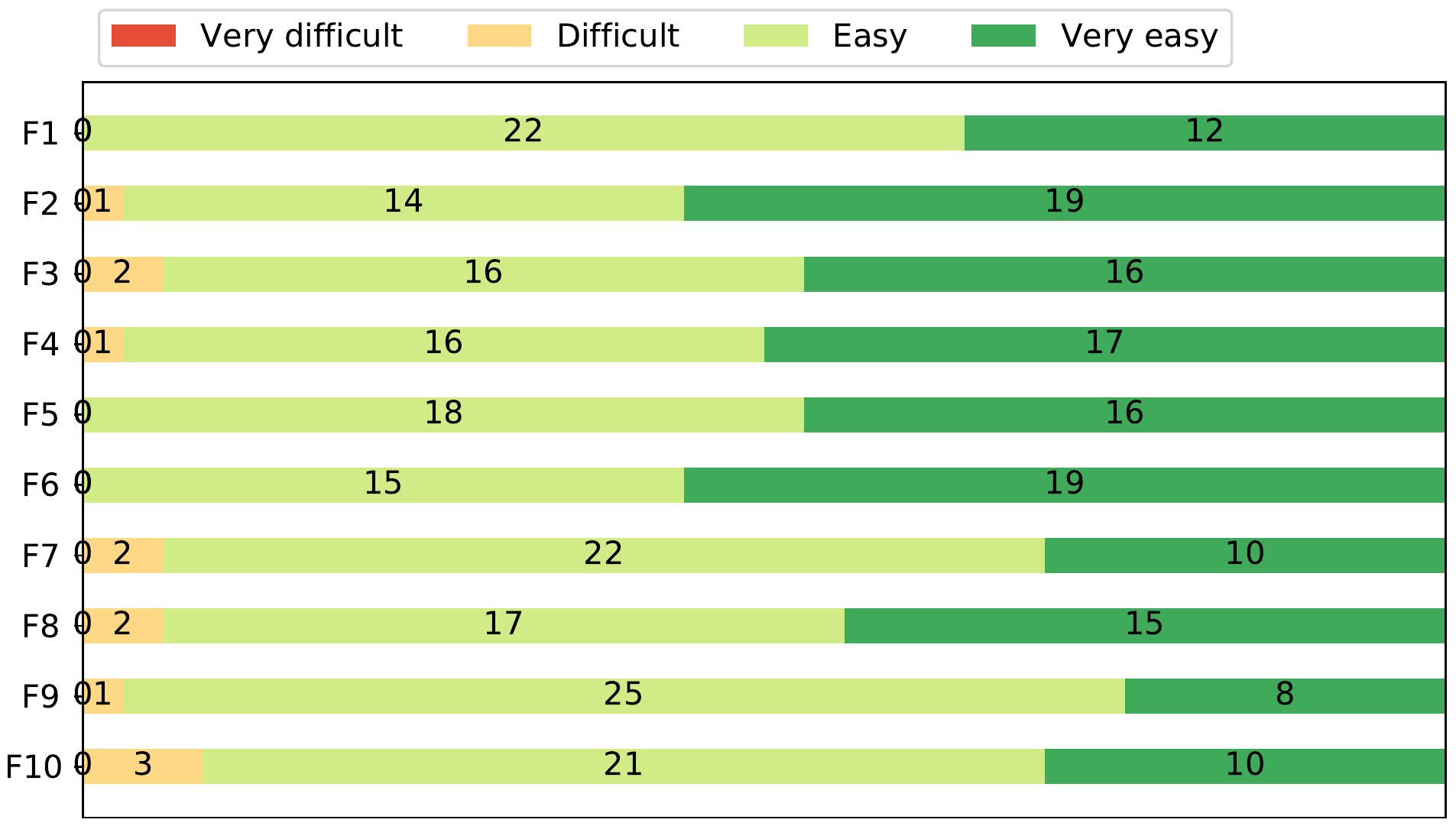}
		\caption{Quantity statistics of readability scores for 10 interfaces' descriptions, which are automatically generated through \textsc{Stan}.}
		\label{fig_rq3_2}
		\vspace{-1ex}
	\end{minipage}
\end{figure*}

\subsection{RQ2 Accuracy}
\label{sec:rq2_accuracy}
\begin{table}[ht!]
	\centering
	\vspace*{1ex}
	\scriptsize
	\caption{Quantity statistics of \textsc{Stan}'s accuracy of tagging insecure bytecodes, FD (functionality description), UD (usage description), and PD (payment description).}
	\vspace{-0ex}
	\label{rq2_1}
	\begin{tabular}{|c|c|c||c|c|c|}
		\hline
		\textbf{DS}&\textbf{Result}&\textbf{Accuracy}&\textbf{DS}&\textbf{Result}&\textbf{Accuracy} \\ \hline
		\textbf{1'}&NF tag&62\ding{51} / 0\ding{55}&\textbf{2'}&UD&4,180\ding{51} / 0\ding{55} \\ \hline
		\textbf{1'}&JE tag&3\ding{51} / 0\ding{55}&\textbf{2'}&PD non-payable&2,546\ding{51}/0\ding{55} \\ \hline
		\textbf{2'}&FD&217\ding{51} / 12\ding{59} / 1\ding{55}&\textbf{2'}&PD payable&1,634\ding{51}/0\ding{55} \\ \hline
	\end{tabular}
	\vspace{2ex}
\end{table}
In this section, we evaluate to what extent can \textsc{Stan} accurately describe bytecodes. We first evaluate the insecure contracts' bytecodes tagged in DS1', and FD/UD/PD in DS2', whose statistics are shown in Table~\ref{rq2_1}. Unlike DS1', all bytecodes in DS2' have corresponding source codes, which makes it possible for us to evaluate the accuracy of their FD/UD/PD/BD through sources' review and static analysis.

By using \textsc{Disasm}~\cite{19}, which is a disassembler tool, we acquire all 62 tagged NF bytecodes' corresponding opcodes and retrieve operation \texttt{PUSH4} in them. No matter which type of runtime function dispatcher, there is bound to exist \texttt{PUSH4} operation. However, \texttt{PUSH4} is not retrieved in the 62 tagged NF contracts. To further validate our conclusion, we analyze all history transactions of these 62 NF contracts. Only 2 of these 62 contracts were invoked after creation. All the 10 history transactions of one contract (Address at Mainnet: \seqsplit{0x84161a5491D9A9348ED48d44b2c717C9ab92B4F3}) were reverted, which wastes users' gas, and the other contract (Address at Mainnet: \seqsplit{0xbB38048902107b62A680db6bA69d6d356D6A8014}) maliciously received user's ETH attached in transaction. For the 3 tagged JE contracts' bytecodes, only 1 contract (Address at Mainnet: \seqsplit{0x9C88d1967fE2653da893B742aDa960D6570592b7}) was invoked after creation, which encountered error \textsl{``Bad jump destination''}. For the other 2 contracts, we re-deploy them in our private local chain and invoke them, the same error was encountered as a result. 

For the FD, we randomly select 20 bytecodes from DS2' and totally get 230 interfaces with their FDs generated by \textsc{Stan}. To avoid the threat of inter-rater reliability, we ask three different people to evaluate their accuracy. Through manual sources' review, we discover that 217 (94.3\%) interfaces' FDs are accurate, while 12 (5.2\%) interfaces' FDs are inaccurate and 1 (0.4\%) interface's FD is wrong. Inaccurate and wrong FDs are mainly due to that there are incomprehensible abbreviations in some FDs. For example, FD of function \texttt{getBlockNM()} is \textsl{``Gets block nm''}. 

To evaluate the accuracy of UD, leveraging \textsc{Scans}, we acquire all of the 4,180 functions' text signatures from their source codes. Then we use Keccak-256 hash algorithm to calculate bytes signature for each of them. Verified by comparison with bytecodes' descriptions, 100\% of the 4,180 functions' text signatures in UD are correct. 

For the PD, we first acquire ABIs of all the bytecodes' corresponding Solidity sources in DS2' from Etherscan. Leveraging \textsc{Scans}, we statically analyze the \texttt{payable} field (\texttt{True} or \texttt{False}) of every external/public function in ABIs. Then we compare the results to the payment descriptions generated through \textsc{Stan}. As a result, 100\% of 2,546 non-payable and 1,634 payable interfaces, which are all described from runtime bytecodes through \textsc{Stan}, are correct. 

We further evaluate \textsc{Stan}'s accuracy of BD (behavior description). \textsc{Stan} totally detects and describes 72 different interfaces with message-call behaviors, whose statistics are shown in Table~\ref{rq2_4}. Leveraging \textsc{Scans}, we statically analyze the AST of those functions' corresponding Solidity sources, trying to detect Solidity statements corresponding to these specific message-call behaviors. For the other described interfaces without message-call behavior, there is no related statement detected. Note that the user-defined contract call behavior has no fixed Solidity statement, and we check the 26 cases manually through source review.

\begin{table}[ht!]
	\centering
	\vspace*{1ex}
	\scriptsize
	\caption{Quantity statistics of \textsc{Stan}'s accuracy of message-call behaviors' description in BD. $\star$ marks pre-compiled contract calls.}
	\vspace{0ex}
	\label{rq2_4}
	\begin{tabular}{|c|c|c|c|}
		\hline
		\textbf{DS}&\textbf{Result}&\textbf{Accuracy}&\textbf{Evaluated statement} \\ \hline
		\textbf{2'}&ETH transfer&20 \ding{51}&{transfer()/call.value()/selfdestruct()} \\ \hline
		\textbf{2'}&$\star$ECDSA sig recovery&16 \ding{51}&{ecrecover(bytes32,uint8,bytes32,...)} \\ \hline
		\textbf{2'}&U-defined contract call&26 \ding{51}&{N/A} \\ \hline
		\textbf{2'}&$\star$SHA-256 hash&2 \ding{51}&{sha256(bytes)} \\ \hline
		\textbf{2'}&$\star$RIPEMD-256 hash&1 \ding{51}&{ripemd160(bytes)} \\ \hline
		\textbf{2'}&Contract deployment&7 \ding{51}&{new CONTRACT} \\ \hline
	\end{tabular}
	\vspace{2ex}
\end{table}

\textbf{Answer to RQ2 (Accuracy):} 100\% of the insecure contracts are correctly tagged, and more than 94.3\% of generated functionality descriptions are correct. 100\% of generated usage/payment/behavior descriptions are correct. \textsc{Stan} can accurately describe bytecodes of smart contracts.

\subsection{RQ3 Readability}
\label{sec:rq3}
\vspace{1ex}
In this section, we evaluate the readability of descriptions generated through \textsc{Stan}. First, from the 4,180 described interfaces in DS2', we \textit{randomly} select 10 interfaces and make sure that they all have annotations written by developers in their corresponding Solidity sources. Then we try to evaluate the readability of these 10 interfaces' descriptions generated through \textsc{Stan} from their bytecodes, and their annotations written by developers. Second, we design a questionnaire through \textsc{Surveymonkey}~\cite{20}. We set a screening question only to accept those who have ever used Ethereum before. Furthermore, we set four readability scores and options (\textit{4:very difficult, 3:difficult, 2:easy, 1:very easy}) in 20 evaluation questions. Also, if the responder thinks the annotations or descriptions difficult to read, we set open questions to input their reasons. 

\begin{figure}[ht]
	\centering
	\vspace*{-1ex}
	\includegraphics[width=3.40in]{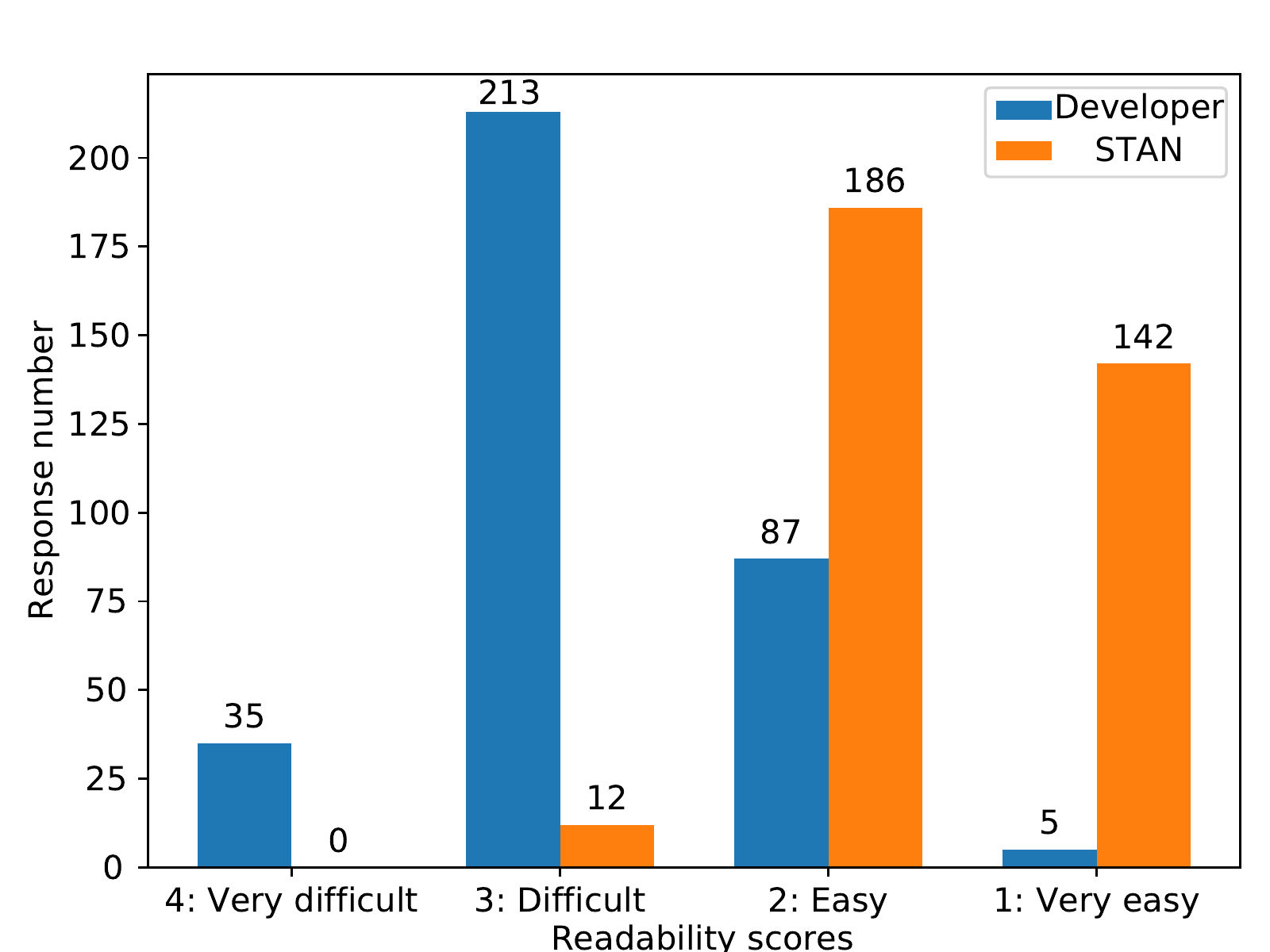}
	\vspace*{-0ex}
	\caption{Quantity statistics of readability scores evaluated manually and their corresponding response number. Note that we totally receive 340 evaluation responses for developer's comments and \textsc{Stan}'s descriptions, respectively.}
	\vspace{2ex}
	\label{fig_rq3_0}
\end{figure}

Third, we publish the questionnaire in BlockFlow~\cite{21}, which is a blockchain development forum. During 4 days, we totally receive 38 responses. However, 2 responses are incomplete, and 2 responses do not meet the screening criteria. Therefore, the completion rate of the questionnaire is 95\% (34/38). Quantity statistics of readability scores and their corresponding response numbers are shown in Figure~\ref{fig_rq3_0}. For total 340 evaluation responses (34 complete questionnaires for 10 interfaces), 72.9\% (248/340) responses think the annotations written by developers are (very) difficult to read, while 96.5\% (328/340) responses think the descriptions generated through \textsc{Stan} are (very) easy to read. 

We also analyze the reasons that responders provide why they think some items (very) difficult to read. We further analyze and compare the readability scores distributions of developers' annotations and \textsc{Stan}'s descriptions, whose quantity statistics are shown in Figure~\ref{fig_rq3_1} and~\ref{fig_rq3_2}. For the descriptions generated through \textsc{Stan}, one responder suggests giving Solidity snippet example to call the bytecodes' interface, which is out of scope of this paper. For the annotations written by developers, there are mainly 3 different reasons. There are 103 responders think the annotations are too simple explanation for interfaces, 36 responders think there exist syntax errors in annotations, and 29 responders think some vocabulary cannot be understood. Through manually checking the corresponding specific content of developers' annotations, function \textit{0x13af4035}'s annotation only has 3 words (\textit{``Change owner address''}), which 21 responders think is too simple. For function \textit{0xa9059cbb}'s annotation (\textit{``Check if the sender has enough. Add the same to the recipient.''}), 16 responders think it has syntax errors. For function \textit{0xa9059cbb}'s annotation (\textit{``SafeMath.sub will throw''}), 12 responders think some vocabulary cannot be understood.
\begin{table*}[ht!]
	\centering
	\vspace*{1ex}
	\scriptsize
	\caption{Statistics of two kinds of non-parametric tests (Kolmogorov-Smirnov Z and Mann-Whitney U) for the readability value of \textsc{Stan}'s descriptions. We compare the bytecodes' descriptions generated through \textsc{Stan}, and annotations in their corresponding Solidity sources written by developers. Note that AVG and STD mean average and standard deviation.}
	\vspace{0ex}
	\label{rq3_1}
	\begin{tabular}{|c|c|c|c|c|c|c|c|}
		\hline
		\tabincell{c}{\textbf{Interface}\\\textbf{ID}}&\tabincell{c}{\textbf{Function}\\\textbf{Bytes signature}}&\tabincell{c}{\textbf{AVG value}\\\textbf{Developer / \textsc{Stan}}}&\tabincell{c}{\textbf{STD value}\\\textbf{Developer / \textsc{Stan}}}&\tabincell{c}{\textbf{p value}\\\textbf{Kolmogorov-Smirnov Z}}&\tabincell{c}{\textbf{h value}\\\textbf{KSZ}}&\tabincell{c}{\textbf{p value}\\\textbf{Mann-Whitney U}}&\tabincell{c}{\textbf{h value}\\\textbf{MWU}} \\ \hline
		1&0xe724529c&2.500 / 1.647&\textbf{0.697} / 0.477&3.118425 * e-05&1&7.142322 * e-07&1 \\ \hline
		2&0x42966c68&2.971 / 1.471&0.514 / 0.554&7.693612 * e-12&1&5.764868 * e-12&1 \\ \hline
		3&0xa9059cbb&2.941 / 1.588&0.338 / 0.589&7.327250 * e-13&1&3.956494 * e-12&1 \\ \hline
		4&0xa9059cbb&2.676 / 1.529&0.468 / 0.543&5.118627 * e-07&1&4.887877 * e-10&1 \\ \hline
		5&0x4bb278f3&2.176 / 1.531&0.381 / 0.499&8.952478 * e-04&1&6.089033 * e-07&1 \\ \hline
		6&0x18160ddd&2.676 / 1.441&0.527 / 0.497&5.118627 * e-07&1&1.067191 * e-10&1 \\ \hline
		7&0xd73dd623&3.118 / 1.764&0.322 / 0.546&1.601234 * e-16&1&2.134156 * e-13&1 \\ \hline
		8&0xb602a917&2.735 / 1.617&\textbf{0.609} / 0.594&8.662292 * e-06&1&8.142437 * e-09&1 \\ \hline
		9&0x74a8f103&\textbf{2.118} / 1.794&0.322 / 0.471&\textbf{3.067583 * e-01}&\textbf{0}&1.129570 * e-03&1 \\ \hline
		10&0x13af4035&2.853 / 1.792&\textbf{0.809} / 0.583&2.204947 * e-06&1&2.625890 * e-07&1 \\ \hline
	\end{tabular}
	\vspace{1ex}
\end{table*}

Leveraging \textsc{Scipy}~\cite{22}, we further analyze the statistical distributions of the readability scores through two kinds of non-parametric tests, whose statistics are shown in Table~\ref{rq3_1}. First, we compare the AVG (average) and STD (standard deviation) values of the readability scores for the bytecodes' descriptions generated through \textsc{Stan} and annotations written by developers. As a result, all of \textsc{Stan}'s descriptions perform better than their corresponding developers' annotations in AVG. Even compared to the best AVG of developer's annotation (i.e., 2.118), \textsc{Stan}'s descriptions are all received better scores. For STD, all of the most significant three values are appeared in developers' annotations. That is to say, the annotations written by different developers, as well as different responders for the same annotations, there exist significant differences. 

Second, we conduct Kolmogorov-Smirnov Z and Mann-Whitney U tests to detect whether the two sets of scores have the same statistical distribution. If the \texttt{p} value is less than 0.05, which is a relatively strict threshold, the result hypotheses value will be 1, and the two sets' statistical distributions are different. As a result, except for interface ID-9's KSZ test, all results show that the two sets of scores have different statistical distributions. Through manual checking, we discover that the annotations of ID-9 perform the best in the 10 samples, which receive scores closest to those of \textsc{Stan}'s descriptions.

\textbf{Answer to RQ3 (Readability):} Compared with the interfaces' annotations written by developers, 96.5\% manual responses think the descriptions generated through \textsc{Stan} are (very) easy to read. Furthermore, \textsc{Stan} can generate more stable and readable descriptions than developers' annotations.

\section{Limitations and Solutions}
\label{sec:discussion}
In this section, we discuss some limitations and the corresponding solutions, which are as follows: 

(1). As described in Section~\ref{sec:expoverview}, we run the symbolic execution engine alone on two datasets to discover that many contracts' bytecodes encounter timeout exception. In future work, we will improve cloud instance's configuration, and use more significant timeout threshold to reduce the number of timeout cases. 

(2). We analyze every execution path and some opcodes’ symbolic values to describe interfaces accurately and comprehensively, with an average analysis time of 87.4s per contract. In future work, we may consider improving \textsc{Stan}'s performance with faster static analysis techniques and evaluating \textsc{Stan} with more comprehensive bytecodes datasets. 

(3). As described in Section~\ref{rq1_adequacy}, some functions' signatures cannot be analyzed perfectly through SWUM. In future work, we will build more and better syntax trees, and add more common word abbreviations in Ethereum to Stanford parser’s rule libraries to improve SWUM analysis. 

(4). \textsc{Stan} can generate four categories of descriptions for each interface, as well as tag two kinds of insecure contracts' bytecodes. In future work, we will conduct more features' and behavior' analysis to improve \textsc{Stan}'s functionalities.

\section{Related Work}
\label{sec:relatedwork}
Loi et al.~\cite{luu2016making} proposed \textsc{Oyente}, which uses symbolic execution to detect security bugs in smart contracts.  Although \textsc{Stan} uses \textsc{Oyente} as its symbolic execution engine,  our analysis of bytecodes is not related to the security bugs studied in~\cite{luu2016making}. There are some other symbolic execution engines for detecting vulnerable in smart contracts~\cite{23}\cite{24}\cite{chang2018scompile}\cite{krupp2018teether}\cite{mueller2018smashing}\cite{nikolic2018finding}\cite{permenevverx}, whose purposes are different from ours. 

Sergei et al.~\cite{tikhomirov2018smartcheck} proposed \textsc{Smartcheck}, a static tool that examines contracts' Solidity sources to detect security bugs. However, it cannot analyze bytecodes directly. There are many other static analysis tools~\cite{25}\cite{brent2018vandal}\cite{grech2018madmax}\cite{tsankov2018security}\cite{tsankov2018securify}\cite{grossman2017online} for detecting different kinds of security issues in smart contracts. Some studies~\cite{hirai2017defining}\cite{sergey2018temporal}\cite{grishchenko2018semantic}\cite{hildenbrandt2018kevm}\cite{amani2018towards} employ formal methods to verify security properties of smart contracts and EVM, whose purposes differ from our paper. 

Matt~\cite{suiche2017porosity} proposed \textsc{Porosity}, a decompiler for contracts' bytecodes. However, there are still many challenges to generate accurate and readable source codes. Similarly, there are some research~\cite{grech2019gigahorse}\cite{brent2018vandal}\cite{26}\cite{zhou2018erays}\cite{albert2018ethir}\cite{27} decompiling contracts' bytecodes into user-defined intermediate languages, to improve the readability of runtime bytecodes and to facilitate the analysis of smart contracts. Some other studies (e.g., gas optimization~\cite{chen2017under}~\cite{chen2020}\cite{chen2018towards}) have different purposes. In summary, these studies provide us with valuable inspiration to conduct the first research of describing contracts' bytecodes.


\section{Conclusion}
\label{sec:conclusion}
In this paper, we propose \textsc{Stan}, which leverages symbolic execution and NLP techniques to describe runtime bytecodes of smart contracts. \textsc{Stan} can generate four categories of descriptions in natural language for every interface of bytecodes deployed in Ethereum. We also develop static tool \textsc{Scans} to facilitate us to construct the database for \textsc{Stan}, and facilitate us to evaluate the generated descriptions. Extensive experiments show that \textsc{Stan} can generate adequate, accurate, and readable descriptions for bytecodes. In future work, we will explore other techniques (e.g., deep learning \cite{deepcodesummarization}) to generate better descriptions.

\section{Acknowledgement}
We thank the anonymous reviewers for their helpful comments. This research is partially supported by the Hong Kong General Research Fund (No. 152193/19E) and the National Natural Science Foundation of China (No. 61872057) and National Key R\&D Program of China (2018YFB0804100).

%
\IEEEpeerreviewmaketitle






\normalem
\bibliographystyle{IEEEtran}
\newpage
\bibliography{ref}

\begin{thebibliography}{10}
\providecommand{\url}[1]{#1}
\csname url@samestyle\endcsname
\providecommand{\newblock}{\relax}
\providecommand{\bibinfo}[2]{#2}
\providecommand{\BIBentrySTDinterwordspacing}{\spaceskip=0pt\relax}
\providecommand{\BIBentryALTinterwordstretchfactor}{4}
\providecommand{\BIBentryALTinterwordspacing}{\spaceskip=\fontdimen2\font plus
\BIBentryALTinterwordstretchfactor\fontdimen3\font minus
  \fontdimen4\font\relax}
\providecommand{\BIBforeignlanguage}[2]{{%
\expandafter\ifx\csname l@#1\endcsname\relax
\typeout{** WARNING: IEEEtran.bst: No hyphenation pattern has been}%
\typeout{** loaded for the language `#1'. Using the pattern for}%
\typeout{** the default language instead.}%
\else
\language=\csname l@#1\endcsname
\fi
#2}}
\providecommand{\BIBdecl}{\relax}
\BIBdecl

\bibitem{yue2016healthcare}
X.~Yue, H.~Wang, D.~Jin, M.~Li, and W.~Jiang, ``Healthcare data gateways: found
  healthcare intelligence on blockchain with novel privacy risk control,'' in
  \emph{Journal of medical systems}, vol.~40, no.~10, 2016.

\bibitem{esposito2018blockchain}
C.~Esposito, A.~De~Santis, G.~Tortora, H.~Chang, and K.-K.~R. Choo,
  ``Blockchain: A panacea for healthcare cloud-based data security and
  privacy?'' in \emph{IEEE Cloud Computing}, vol.~5, no.~1, 2018, pp. 31--37.

\bibitem{samaniego2016blockchain}
M.~Samaniego and R.~Deters, ``Blockchain as a service for iot,'' in
  \emph{Proceedings of the IEEE International Conference on Internet of Things
  and IEEE Green Computing and Communications and IEEE Cyber, Physical and
  Social Computing and IEEE Smart Data}.\hskip 1em plus 0.5em minus 0.4em\relax
  IEEE, 2016, pp. 433--436.

\bibitem{liang2017provchain}
X.~Liang, S.~Shetty, D.~Tosh, C.~Kamhoua, K.~Kwiat, and L.~Njilla, ``Provchain:
  A blockchain-based data provenance architecture in cloud environment with
  enhanced privacy and availability,'' in \emph{Proceedings of the 17th
  international symposium on cluster, cloud and grid computing}.\hskip 1em plus
  0.5em minus 0.4em\relax IEEE/ACM, 2017, pp. 468--477.

\bibitem{zheng2018blockchain}
Z.~Zheng, S.~Xie, H.-N. Dai, X.~Chen, and H.~Wang, ``Blockchain challenges and
  opportunities: A survey,'' in \emph{International Journal of Web and Grid
  Services}, vol.~14, no.~4, 2018, pp. 352--375.

\bibitem{2}
\BIBentryALTinterwordspacing
Ethereum, ``Ethereum transaction chart,'' 2020. [Online]. Available:
  \url{https://etherscan.io/chart/tx}
\BIBentrySTDinterwordspacing

\bibitem{1_2}
\BIBentryALTinterwordspacing
Ethereum-community, ``Contracts with verified source codes,'' 2020. [Online].
  Available: \url{https://etherscan.io/contractsVerified}
\BIBentrySTDinterwordspacing

\bibitem{grech2019gigahorse}
N.~Grech, L.~Brent, B.~Scholz, and Y.~Smaragdakis, ``Gigahorse: thorough,
  declarative decompilation of smart contracts,'' in \emph{Proceedings of the
  International Conference on Software Engineering}.\hskip 1em plus 0.5em minus
  0.4em\relax IEEE, 2019, pp. 1176--1186.

\bibitem{zhou2018erays}
Y.~Zhou, D.~Kumar, S.~Bakshi, J.~Mason, A.~Miller, and M.~Bailey, ``Erays:
  reverse engineering ethereum's opaque smart contracts,'' in \emph{Proceedings
  of the 27th USENIX Security Symposium}.\hskip 1em plus 0.5em minus
  0.4em\relax USENIX, 2018, pp. 1371--1385.

\bibitem{7}
\BIBentryALTinterwordspacing
Ethereum, ``The yellow paper: Ethereum's formal specification,'' 2019.
  [Online]. Available: \url{https://github.com/ethereum/yellowpaper}
\BIBentrySTDinterwordspacing

\bibitem{1}
\BIBentryALTinterwordspacing
Ethereum-community, ``Etherscan,'' 2020. [Online]. Available:
  \url{https://etherscan.io}
\BIBentrySTDinterwordspacing

\bibitem{suiche2017porosity}
M.~Suiche, ``Porosity: A decompiler for blockchain-based smart contracts
  bytecode,'' in \emph{Defcon}, vol.~25, 2017, p.~11.

\bibitem{luu2016making}
L.~Luu, D.-H. Chu, H.~Olickel, P.~Saxena, and A.~Hobor, ``Making smart
  contracts smarter,'' in \emph{Proceedings of the SIGSAC Conference on
  Computer and Communications Security}.\hskip 1em plus 0.5em minus 0.4em\relax
  ACM, 2016, pp. 254--269.

\bibitem{23}
\BIBentryALTinterwordspacing
C.~Inc, ``Mythril,'' 2019. [Online]. Available:
  \url{https://github.com/ConsenSys/mythril}
\BIBentrySTDinterwordspacing

\bibitem{tikhomirov2018smartcheck}
S.~Tikhomirov, E.~Voskresenskaya, I.~Ivanitskiy, R.~Takhaviev, E.~Marchenko,
  and Y.~Alexandrov, ``Smartcheck: Static analysis of ethereum smart
  contracts,'' in \emph{Proceedings of the 1st International Workshop on
  Emerging Trends in Software Engineering for Blockchain}.\hskip 1em plus 0.5em
  minus 0.4em\relax IEEE, 2018, pp. 9--16.

\bibitem{25}
J.~Feist, G.~Grieco, and A.~Groce, ``journal: a static analysis framework for
  smart contracts,'' in \emph{Proceedings of the 2nd International Workshop on
  Emerging Trends in Software Engineering for Blockchain}.\hskip 1em plus 0.5em
  minus 0.4em\relax IEEE, 2019, pp. 8--15.

\bibitem{CFI}
T.~Chen, Z.~Li, Y.~Zhang, X.~Luo, T.~Wang, T.~Hu, X.~Xiao, D.~Wang, J.~Huang,
  and X.~Zhang, ``A large-scale empirical study on control flow identification
  of smart contracts,'' in \emph{Proceedings of International Symposium on
  Empirical Software Engineering and Measurement}, 2019.

\bibitem{TokenScope}
T.~Chen, Y.~Zhang, Z.~Li, X.~Luo, T.~Wang, R.~Cao, X.~Xiao, and X.~Zhang,
  ``Tokenscope: Automatically detecting inconsistent behaviors of
  cryptocurrency tokens in ethereum,'' in \emph{Proceedings of the SIGSAC
  Conference on Computer and Communications Security}.\hskip 1em plus 0.5em
  minus 0.4em\relax ACM, 2019.

\bibitem{hirai2017defining}
Y.~Hirai, ``Defining the ethereum virtual machine for interactive theorem
  provers,'' in \emph{Proceedings of the International Conference on Financial
  Cryptography and Data Security}.\hskip 1em plus 0.5em minus 0.4em\relax
  Springer, 2017, pp. 520--535.

\bibitem{sergey2018temporal}
I.~Sergey, A.~Kumar, and A.~Hobor, ``Temporal properties of smart contracts,''
  in \emph{Proceedings of the International Symposium on Leveraging
  Applications of Formal Methods}.\hskip 1em plus 0.5em minus 0.4em\relax
  Springer, 2018, pp. 323--338.

\bibitem{wan2019discussed}
Z.~Wan, X.~Xia, and A.~E. Hassan, ``What is discussed about blockchain? a case
  study on the use of balanced lda and the reference architecture of a domain
  to capture online discussions about blockchain platforms across the stack
  exchange communities,'' in \emph{IEEE Transactions on Software Engineering},
  2019.

\bibitem{chen2018understanding}
T.~Chen, Y.~Zhu, Z.~Li, J.~Chen, X.~Li, X.~Luo, X.~Lin, and X.~Zhange,
  ``Understanding ethereum via graph analysis,'' in \emph{Proceedings of the
  IEEE Conference on Computer Communications}.\hskip 1em plus 0.5em minus
  0.4em\relax IEEE, 2018.

\bibitem{zou2019}
W.~Zou, D.~Lo, P.~S. Kochhar, X.-B.~D. Le, X.~Xia, Y.~Feng, Z.~Chen, and B.~Xu,
  ``Smart contract development: Challenges and opportunities,'' in \emph{IEEE
  Transactions on Software Engineering}, 2019.

\bibitem{chen2020understanding}
T.~Chen, Z.~Li, Y.~Zhu, J.~Chen, X.~Luo, J.~C.-S. Lui, X.~Lin, and X.~Zhang,
  ``Understanding ethereum via graph analysis,'' \emph{ACM Transactions on
  Internet Technology (TOIT)}, vol.~20, no.~2, 2020.

\bibitem{chen2017adaptive}
T.~Chen, X.~Li, Y.~Wang, J.~Chen, Z.~Li, X.~Luo, M.~H. Au, and X.~Zhang, ``An
  adaptive gas cost mechanism for ethereum to defend against under-priced dos
  attacks,'' in \emph{Proceedings of the International Conference on
  Information Security Practice and Experience}.\hskip 1em plus 0.5em minus
  0.4em\relax Springer, 2017.

\bibitem{li2020survey}
X.~Li, P.~Jiang, T.~Chen, X.~Luo, and Q.~Wen, ``A survey on the security of
  blockchain systems,'' in \emph{Future Generation Computer Systems}, vol.
  107.\hskip 1em plus 0.5em minus 0.4em\relax Elsevier, 2020, pp. 841--853.

\bibitem{11}
\BIBentryALTinterwordspacing
Ethereum, ``Ethereum natural specification format,'' 2019. [Online]. Available:
  \url{https://github.com/ethereum/wiki/wiki/Ethereum-Natural-Specification-Format}
\BIBentrySTDinterwordspacing

\bibitem{10}
\BIBentryALTinterwordspacing
Ethereum-community, ``Ethereum improvement proposals,'' 2019. [Online].
  Available: \url{https://eips.ethereum.org/erc}
\BIBentrySTDinterwordspacing

\bibitem{31}
\BIBentryALTinterwordspacing
Ethereum, ``Erc20 tokens,'' 2020. [Online]. Available:
  \url{https://etherscan.io/tokens}
\BIBentrySTDinterwordspacing

\bibitem{mihalcea2004textrank}
R.~Mihalcea and P.~Tarau, ``Textrank: Bringing order into text,'' in
  \emph{Proceedings of the conference on empirical methods in natural language
  processing}, 2004.

\bibitem{broder1997resemblance}
A.~Z. Broder, ``On the resemblance and containment of documents,'' in
  \emph{Proceedings of Compression and Complexity of Sequences}, 1997.

\bibitem{sridhara2010towards}
G.~Sridhara, E.~Hill, D.~Muppaneni, L.~Pollock, and K.~Vijay-Shanker, ``Towards
  automatically generating summary comments for java methods,'' in
  \emph{Proceedings of the international conference on automated software
  engineering}.\hskip 1em plus 0.5em minus 0.4em\relax IEEE/ACM, 2010, pp.
  43--52.

\bibitem{wordninja}
\BIBentryALTinterwordspacing
D.~Anderson, ``Wordninja,'' 2019. [Online]. Available:
  \url{https://github.com/keredson/wordninja}
\BIBentrySTDinterwordspacing

\bibitem{13}
J.~Nivre, M.-C. De~Marneffe, F.~Ginter, Y.~Goldberg, J.~Hajic, C.~D. Manning,
  R.~McDonald, S.~Petrov, S.~Pyysalo, N.~Silveira \emph{et~al.}, ``Universal
  dependencies v1: A multilingual treebank collection,'' in \emph{Proceedings
  of the Tenth International Conference on Language Resources and Evaluation},
  2016, pp. 1659--1666.

\bibitem{3}
\BIBentryALTinterwordspacing
EFSD, ``Ethereum function signature database,'' 2020. [Online]. Available:
  \url{https://www.4byte.directory/}
\BIBentrySTDinterwordspacing

\bibitem{14}
K.~Chodorow, \emph{MongoDB: the definitive guide: powerful and scalable data
  storage}.\hskip 1em plus 0.5em minus 0.4em\relax O'Reilly Media Inc, 2013.

\bibitem{29}
\BIBentryALTinterwordspacing
Ethereum, ``Solidity,'' 2020. [Online]. Available:
  \url{https://solidity.readthedocs.io}
\BIBentrySTDinterwordspacing

\bibitem{17}
A.~Gatt and E.~Reiter, ``Simplenlg: A realisation engine for practical
  applications,'' in \emph{Proceedings of the 12th European Workshop on Natural
  Language Generation}, 2009, pp. 90--93.

\bibitem{18}
\BIBentryALTinterwordspacing
S.~Myint, ``Language-check,'' 2019. [Online]. Available:
  \url{https://github.com/myint/language-check}
\BIBentrySTDinterwordspacing

\bibitem{19}
\BIBentryALTinterwordspacing
Ethereum, ``Evm disassembler,'' 2019. [Online]. Available:
  \url{https://github.com/Arachnid/evmdis}
\BIBentrySTDinterwordspacing

\bibitem{20}
\BIBentryALTinterwordspacing
S.~Inc, ``Surveymonkey,'' 2020. [Online]. Available:
  \url{https://www.surveymonkey.com}
\BIBentrySTDinterwordspacing

\bibitem{21}
\BIBentryALTinterwordspacing
B.~Community, ``Blockflow,'' 2020. [Online]. Available:
  \url{https://blockflow.net/}
\BIBentrySTDinterwordspacing

\bibitem{22}
E.~Bressert, \emph{SciPy and NumPy: an overview for developers}.\hskip 1em plus
  0.5em minus 0.4em\relax O'Reilly Media Inc., 2012.

\bibitem{24}
\BIBentryALTinterwordspacing
T.~of~Bits, ``Manticore,'' 2019. [Online]. Available:
  \url{https://github.com/trailofbits/manticore}
\BIBentrySTDinterwordspacing

\bibitem{chang2018scompile}
J.~Chang, B.~Gao, H.~Xiao, J.~Sun, and Z.~Yang, ``scompile: Critical path
  identification and analysis for smart contracts,'' in \emph{preprint
  arXiv:1808.00624}, 2018.

\bibitem{krupp2018teether}
J.~Krupp and C.~Rossow, ``teether: Gnawing at ethereum to automatically exploit
  smart contracts,'' in \emph{Proceedings of the 27th USENIX Security
  Symposium}.\hskip 1em plus 0.5em minus 0.4em\relax USENIX, 2018, pp.
  1317--1333.

\bibitem{mueller2018smashing}
B.~Mueller, ``Smashing ethereum smart contracts for fun and real profit,'' in
  \emph{Proceedings of the 9th Annual HITB Security Conference}, 2018.

\bibitem{nikolic2018finding}
I.~Nikoli{\'c}, A.~Kolluri, I.~Sergey, P.~Saxena, and A.~Hobor, ``Finding the
  greedy, prodigal, and suicidal contracts at scale,'' in \emph{Proceedings of
  the 34th Annual Computer Security Applications Conference}.\hskip 1em plus
  0.5em minus 0.4em\relax ACM, 2018.

\bibitem{permenevverx}
A.~Permenev, D.~Dimitrov, P.~Tsankov, D.~Drachsler-Cohen, and M.~Vechev,
  ``Verx: Safety verification of smart contracts,'' in \emph{Proceedings of the
  41st Symposium on Security and Privacy}.\hskip 1em plus 0.5em minus
  0.4em\relax IEEE, 2020.

\bibitem{brent2018vandal}
L.~Brent, A.~Jurisevic, M.~Kong, E.~Liu, F.~Gauthier, V.~Gramoli, R.~Holz, and
  B.~Scholz, ``Vandal: A scalable security analysis framework for smart
  contracts,'' in \emph{preprint arXiv:1809.03981}, 2018.

\bibitem{grech2018madmax}
N.~Grech, M.~Kong, A.~Jurisevic, L.~Brent, B.~Scholz, and Y.~Smaragdakis,
  ``Madmax: Surviving out-of-gas conditions in ethereum smart contracts,'' in
  \emph{Proceedings of the ACM on Programming Languages}, vol.~2.\hskip 1em
  plus 0.5em minus 0.4em\relax ACM, 2018, p. 116.

\bibitem{tsankov2018security}
P.~Tsankov, ``Security analysis of smart contracts in datalog,'' in
  \emph{Proceedings of the International Symposium on Leveraging Applications
  of Formal Methods}.\hskip 1em plus 0.5em minus 0.4em\relax Springer, 2018,
  pp. 316--322.

\bibitem{tsankov2018securify}
P.~Tsankov, A.~Dan, D.~Drachsler-Cohen, A.~Gervais, F.~Buenzli, and M.~Vechev,
  ``Securify: Practical security analysis of smart contracts,'' in
  \emph{Proceedings of the SIGSAC Conference on Computer and Communications
  Security}.\hskip 1em plus 0.5em minus 0.4em\relax ACM, 2018, pp. 67--82.

\bibitem{grossman2017online}
S.~Grossman, I.~Abraham, G.~Golan-Gueta, Y.~Michalevsky, N.~Rinetzky, M.~Sagiv,
  and Y.~Zohar, ``Online detection of effectively callback free objects with
  applications to smart contracts,'' in \emph{Proceedings of the ACM on
  Programming Languages}.\hskip 1em plus 0.5em minus 0.4em\relax ACM, 2017,
  p.~48.

\bibitem{grishchenko2018semantic}
I.~Grishchenko, M.~Maffei, and C.~Schneidewind, ``A semantic framework for the
  security analysis of ethereum smart contracts,'' in \emph{Proceedings of the
  International Conference on Principles of Security and Trust}.\hskip 1em plus
  0.5em minus 0.4em\relax Springer, 2018, pp. 243--269.

\bibitem{hildenbrandt2018kevm}
E.~Hildenbrandt, M.~Saxena, N.~Rodrigues, X.~Zhu, P.~Daian, D.~Guth, B.~Moore,
  D.~Park, Y.~Zhang, A.~Stefanescu \emph{et~al.}, ``Kevm: A complete formal
  semantics of the ethereum virtual machine,'' in \emph{Proceedings of the 31st
  Computer Security Foundations Symposium}.\hskip 1em plus 0.5em minus
  0.4em\relax IEEE, 2018.

\bibitem{amani2018towards}
S.~Amani, M.~B{\'e}gel, M.~Bortin, and M.~Staples, ``Towards verifying ethereum
  smart contract bytecode in isabelle/hol,'' in \emph{Proceedings of the 7th
  SIGPLAN International Conference on Certified Programs and Proofs}.\hskip 1em
  plus 0.5em minus 0.4em\relax ACM, 2018, pp. 66--77.

\bibitem{26}
\BIBentryALTinterwordspacing
P.~Software, ``Jeb,'' 2019. [Online]. Available:
  \url{https://www.pnfsoftware.com/jeb/evm}
\BIBentrySTDinterwordspacing

\bibitem{albert2018ethir}
E.~Albert, P.~Gordillo, B.~Livshits, A.~Rubio, and I.~Sergey, ``Ethir: A
  framework for high-level analysis of ethereum bytecode,'' in
  \emph{Proceedings of the International Symposium on Automated Technology for
  Verification and Analysis}.\hskip 1em plus 0.5em minus 0.4em\relax Springer,
  2018, pp. 513--520.

\bibitem{27}
R.~Stortz, ``Rattle: an ethereum evm binary analysis framework,'' in
  \emph{REcon Montreal}, 2019.

\bibitem{chen2017under}
T.~Chen, X.~Li, X.~Luo, and X.~Zhang, ``Under-optimized smart contracts devour
  your money,'' in \emph{Proceedings of the 24th International Conference on
  Software Analysis, Evolution and Reengineering}.\hskip 1em plus 0.5em minus
  0.4em\relax IEEE, 2017.

\bibitem{chen2020}
T.~Chen, Y.~Feng, Z.~Li, H.~Zhou, X.~Luo, X.~Li, X.~Xiao, J.~Chen, and
  X.~Zhang, ``Gaschecker: Scalable analysis for discovering gas-inefficient
  smart contracts,'' in \emph{IEEE Transactions on Emerging Topics in
  Computing}.\hskip 1em plus 0.5em minus 0.4em\relax IEEE, 2020.

\bibitem{chen2018towards}
T.~Chen, Z.~Li, H.~Zhou, J.~Chen, X.~Luo, X.~Li, and X.~Zhang, ``Towards saving
  money in using smart contracts,'' in \emph{Proceedings of the 40th
  International Conference on Software Engineering}.\hskip 1em plus 0.5em minus
  0.4em\relax IEEE, 2018.

\bibitem{deepcodesummarization}
J.~Zhang, X.~Wang, H.~Zhang, H.~Sun, and X.~Liu, ``Retrieval-based neural
  source code summarization,'' in \emph{Proceedings of the International
  Conference on Software Engineering}.\hskip 1em plus 0.5em minus 0.4em\relax
  IEEE, 2020.

\end{thebibliography}
\end{document}